\documentstyle[12pt,eqsecnum,aps,fleqn,prb]{revtex}

\begin{document}

\title{The basis of nonlocal curvature invariants in quantum gravity
theory.
(Third order.)}
\vspace{10mm}
\author{\\
  A. O. Barvinsky\hbox{$^{1,\rm a)}$},
  Yu. V. Gusev\hbox{$^{1,\rm b)}$},
  G. A. Vilkovisky\hbox{$^{2}$} \\and\\
  V. V. Zhytnikov\hbox{$^{1,\rm c)}$}
 }

\maketitle
\begin{center}
$^{1}$ {Nuclear Safety Institute, Bolshaya Tulskaya 52, Moscow
113191,
Russia}\\$^{2}${Lebedev Physics Institute, Leninsky Prospect 53,
Moscow 117924,
Russia}
\end{center}
\vspace{20mm}

\begin{abstract}
A complete basis of nonlocal invariants in quantum gravity theory is
built to
third order in spacetime curvature and matter-field strengths. The
nonlocal
identities are obtained which reduce this basis for manifolds with
dimensionality $2\omega<6$. The present results are used in
heat-kernel theory,
theory of gauge fields and serve as a basis for the model-independent
approach
to quantum gravity and, in particular, for the study of nonlocal
vacuum effects
in the gravitational collapse problem.
\end{abstract}

\vspace{10mm}

\thispagestyle{empty}
\footnoterule
\begin{itemize}
\item[$^{\rm a)}$]On leave at the Department
of Physics, University of Alberta, Edmonton T6G 2J1, Canada
\item[$^{\rm b)}$]
On leave at the Department of Physics,
University of Manitoba,
Winnipeg R3T 2N2, Canada
\item[$^{\rm c)}$]
On leave at the Department of Physics of
National Central University,
Chung-li, Taiwan 320
\end{itemize}

\pagebreak

\section{Introduction}
\indent
The concept of nonlocal invariant of a given order in the curvature
(or, more
generally, in the field strength) comes from quantum theory of gauge
fields.
The effective action of quantum operator fields is a nonlocal
functional of the
respective c-number fields \cite{DW63,JLas,Hooft-Hon,DDI,Gosp} which,
for given
quantum states, represent the matrix elements or expectation values
\cite{in,Frolov,Hartle}. For certain quantum states this functional
can be
expanded in powers of the deviation of the field argument from its
value in a
flat and emty spacetime, and the coefficients of this expansion are
exact
vertex functions of the theory. In the case of gauge fields, the
effective
action is an invariant functional, and the calculations can be done
so that
each term of the expansion is invariant
\cite{PRep,Frfest,I,II,III,IV}. The
expansion is then in powers of field strengths (curvatures) but its
coefficients ( the nonlocal form factors) are also field dependent.
The meaning
of such an expansion is that the effective action is obtained with
accuracy
$O\,[\,\Re^N\,]$ i.e. {\it up to} $N$-th power in $\Re$ where $\Re$
is the
collective notation for the full set of the field strengths of the
theory. Each
term of the expansion contains $N$ curvatures $\Re$ explicitly and is
defined
up to $O\,[\,\Re^{N+1}\,]$. By definition, such term is a nonlocal
invariant of
order $N$ in the curvature.

All invariants of order $N$ form a linear space. One still needs a
specification of the class of nonlocal form factors which serve as
coefficients
of linear combinations in this space. Quantum field theory picks up a
quite
definite class which we consider below. With these specifications, we
describe
the general procedure of building the basis of nonlocal invariants of
$N$-th
order and build it explicitly to third order in the curvature
inclusive.

The results obtained in this way are used in the model-independent
approach to
quantum gravity \cite{CQG,Armen}. The central object of this approach
is the
effective action for the in-vacuum state \cite{DW83} which is built
by adopting
the kinematic rules of quantum field theory. It is then a nonlocal
functional
of the class considered here. Owing to the properties of the state
\cite{CQG},
it can be expanded in a basis of $N$-th order curvature invariants
but the
nonlocal form factors in this expansion are left unspecified. These
form
factors stand for the dynamical information contained in an unknown
model of
the vacuum. The purpose of the approach is to relate the properties
of the
spacetime solving the expectation-value equations to the properties
of the
vacuum form factors. Thus, it has been shown that the requirement of
the
asymptotic flatness of the solution leads to quite definite
asymptotic
behaviours of the form factors \cite{Armen}. The physical problem
which is
considered within this approach and for which the nonlocal vacuum
effects play
an important role is the gravitational collapse problem. The Hawking
radiation
\cite{Hawking} and its backreaction on the metric is one such effect
\cite{Frolov}. It has been shown \cite{GAV-Strasb,Armen} that in four
dimensions this effect starts with cubic in the curvature terms of
the vacuum
action. For this reason the explicit construction of the basis in the
present
paper extends to the third order.

Another application that the present technique finds is in heat
kernel theory.
One needs the heat kernel as an intermidiate stage in calculating the
effective
action in quantum field theory \cite{Schw2,DW63,PRep}, but heat
kernel  is, of
course, of self value. The heat kernel is a nonlocal functional of
the
background fields contained in its Hamiltonian operator, and it can
be
calculated as an expansion in nonlocal invariants of $N$-th order
\cite{I,II,IV,Zeln,GAV-Strasb,JMP}.

In the case of the in-vacuum state, the effective equations both in
expectation-value theory and transition-amplitude theory are obtained
by
certain rules from the effective action of euclidean theory
\cite{Frolov,II}
(see also Ref.\onlinecite{Hartle}). Therefore, in the present paper
we confine
ourselves to the consideration of asymptotically flat euclidean
spaces.

The basis of nonlocal invariants is built for an arbitrary dimension
$2\omega$
but, for low-dimensional manifolds, it is redundant because there
exist hidden
identities between the nonlocal invariants of a given order. Most of
the
present paper is in fact devoted to a systematic analysis of these
identities.
In the case of cubic invariants, the redundancy takes place for
$2\omega<6$.
For $2\omega=4$, there exists a single nontrivial identity reducing
the basis
of purely gravitational cubic invariants. As a by-product, we
discover a
mechanism by which the Gauss-Bonnet invariant becomes topological in
four
dimensions.

The paper is organized as follows. In Sect.II we use the Bianchi
identity to
eliminate the Riemann tensor in terms of the Ricci curvature. In
Sects.III and
IV we formulate the notion of nonlocal curvature invariants with the
Ricci
tensor and build the basis of such invariants to third order. Sect.V
contains
the construction of matter invariants and presents the full table of
nonlocal
invariants to third order in curvatures -- gravitational and
matter-field
strengths. Finally, in Sect.VI we derive the hidden identities
between local
and nonlocal cubic invariants in low-dimensional manifolds.

\section{Elimination of the Riemann tensor}
\indent
An important feature of nonlocal curvature invariants in
asymptotically flat
spacetime is that their purely gravitational strength boils down to
the Ricci
tensor, because the Riemann tensor can always be eliminated via the
known
corollary of the Bianchi identity \cite{II}
\mathindent=0pt
\arraycolsep=0pt
\begin{eqnarray}
&&\Box R^{\alpha\beta\mu\nu}\equiv
\frac12\Big(\nabla^\mu \nabla^\alpha R^{\nu\beta}
+\nabla^\alpha \nabla^\mu R^{\nu\beta}
-\nabla^\nu \nabla^\alpha R^{\mu\beta}
-\nabla^\alpha \nabla^\nu R^{\mu\beta}
-\nabla^\mu \nabla^\beta R^{\nu\alpha}
-\nabla^\beta \nabla^\mu R^{\nu\alpha}
\nonumber\\
&& \qquad\qquad
+\nabla^\nu \nabla^\beta R^{\mu\alpha}
+\nabla^\beta \nabla^\nu R^{\mu\alpha}
\Big)
\nonumber\\&& \qquad\qquad
+ R^{[\mu}_{\lambda}
  R^{\nu] \lambda\beta\alpha}
- R^{[\alpha}_{\lambda}
  R^{\beta] \lambda\mu\nu}
-4  R^{\alpha\ [\mu}_{\  \sigma \ \ \lambda}
  R^{\nu] \lambda\beta\sigma}
-  R^{\alpha\beta}_{\ \ \ \sigma\lambda}
  R^{\mu\nu\sigma\lambda},                         \label{eqn:2.1}
\end{eqnarray}
\arraycolsep=3pt
\mathindent=\leftmargini
which can be iteratively solved for $R^{\alpha\beta\mu\nu}$ in terms
of
$R^{\mu\nu}$. In this equation the Ricci tensor plays the role of a
source
which determines the Riemann tensor up to initial or boundary
conditions for
the operator $\Box$. In the case of positive-signature asymptotically
flat
spaces, which guarantee that a Ricci-flat space is flat, the
iterational
solution is uniquely determined  by the Green function $1/\Box$ with
zero
boundary conditions at infinity. Up to the second order in
$R^{\mu\nu}$,
necessary for the construction of the third order nonlocal invariants
below,
this solution is as follows
\mathindent=0pt
\arraycolsep=0pt
\begin{eqnarray}&& R^{\alpha\beta\mu\nu}=
\frac12\,\frac1{\Box}\Big(
 \nabla^\mu \nabla^\alpha R^{\nu\beta}
+\nabla^\alpha \nabla^\mu R^{\nu\beta}
-\nabla^\nu \nabla^\alpha R^{\mu\beta}
-\nabla^\alpha \nabla^\nu R^{\mu\beta}
-\nabla^\mu \nabla^\beta R^{\nu\alpha}
\nonumber\\
&&
\qquad\quad
-\nabla^\beta \nabla^\mu R^{\nu\alpha}
+\nabla^\nu \nabla^\beta R^{\mu\alpha}
+\nabla^\beta \nabla^\nu R^{\mu\alpha}
\Big)
\nonumber\\
&&
\qquad\quad
+\frac1{\Box}\left\{2R^{[\mu}_{\lambda}
(\nabla^{\lambda}
\nabla^{[\alpha}\frac1{\Box} R^{\beta]\nu]}
)
-2R^{[\mu}_{\lambda}(
\nabla^{\nu]}\nabla^{[\alpha}\frac1{\Box} R^{\beta]\lambda})\right.
\nonumber\\
&&
\qquad\quad
-4(\nabla^{\lambda}\nabla^{[\alpha}\frac1{\Box}
R^{\beta]}_{\sigma})
(\nabla_{\lambda}
\nabla^{[\mu}\frac1{\Box} R^{\nu]\sigma})
-8(\nabla^{\lambda}\nabla^{[\alpha}\frac1{\Box}
R^{[\mu}_{\sigma})
(\nabla_{\lambda}\nabla^{\beta]}\frac1{\Box}
R^{\nu]\sigma})
\nonumber\\
&&
\qquad\quad
+4(\nabla_{\lambda}\nabla^{[\alpha}
\frac1{\Box} R^{\beta]}_{\sigma})
(\nabla^{\sigma}\nabla^{[\mu}
\frac1{\Box} R^{\nu]\lambda})
-4(\nabla^{\sigma}\nabla^{[\mu}\frac1{\Box}
R^{[\alpha}_{\lambda})
(\nabla^{\lambda}\nabla^{\beta]}\frac1{\Box}
R^{\nu]}_{\sigma})
\nonumber\\
&&
\qquad\quad
+8(\nabla^{\lambda}\nabla^{\sigma}
\frac1{\Box} R^{[\mu[\alpha})
(\nabla_{\lambda}\nabla^{\beta]}
\frac1{\Box} R^{\nu]}_{\sigma})
+8(\nabla_{\lambda}\nabla^{[\alpha}
\frac1{\Box} R^{[\mu}_{\sigma})
(\nabla^{\nu]}\nabla^{\beta]}
\frac1{\Box} R^{\lambda\sigma})
\nonumber\\
&&
\qquad\quad
-4(\nabla_{\lambda}
\nabla_{\sigma}\frac1{\Box} R^{[\mu[\alpha})
(\nabla^{\nu]}
\nabla^{\beta]}\frac1{\Box} R^{\lambda\sigma})
-4(\nabla_{\lambda}
\nabla_{\sigma}\frac1{\Box} R^{[\mu[\alpha})
(\nabla^{\lambda}
\nabla^{\sigma}\frac1{\Box} R^{\nu]\beta]})
\nonumber\\
&&
\qquad\quad
\left.-4(\nabla^{[\mu}
\nabla^{[\alpha}\frac1{\Box} R_{\lambda\sigma})
(\nabla^{\nu]}
\nabla^{\beta]}\frac1{\Box} R^{\lambda\sigma})
+\Big[\,(\alpha\beta)\leftrightarrow (\mu\nu)\,\Big]\,
\right\}
+{\rm O}[R^3_{..}],                                   \label{eqn:2.2}
\end{eqnarray}
\arraycolsep=3pt
\mathindent=\leftmargini
where the antisymmetrizations on the right-hand side are with respect
to
$\mu\nu$ and $\alpha\beta$ and $(\alpha\beta)\leftrightarrow
(\mu\nu)$ denote
the terms with the obvious permutation of pairs of indices which
reproduces the
symmetries of the Riemann tensor. Here we use a simplified notation
for the
Green functions introduced in \cite{II}. For example, the quantity
	\begin{equation}
	X^{\mu\nu}=\frac{1}{\Box}\,R^{\mu\nu}         \label{eqn:2.3}
	\end{equation}
denotes the solution of the tensor equation  $\Box
\,X^{\mu\nu}=R^{\mu\nu}$
subject to zero boundary conditions at the infinity of the Euclidean
asymptotically flat spacetime. The same convention applies also to
other tensor
quantities like $\nabla^\mu \nabla^\alpha R^{\nu\beta}$, so that the
tensor
properties of the inverse box operator being determined by the tensor
properties of the quantity which it acts on.

Note, that in case of the Lorentzian asymptotically flat spacetime
Ricci-flat
geometry is also flat, provided there is no incoming gravitational
wave
\cite{CQG}, and the equation of the above type also holds with
$1/\Box$
replaced by the retarded Green's function $1/\Box^{\rm ret}$.

\section{Nonlocal invariants of $n$-th order with the Ricci tensor}
\indent
We shall start with the construction of purely gravitational
invariants. In
this case, since the Riemann tensor is eliminated, we are left with
the Ricci
tensor. A regular procedure of building the basis of nonlocal
invariants
referred to in the Introduction looks as follows.

To begin with, consider the diagrams of effective action in quantum
field
theory and assume (for the moment) that all propagators in these
diagrams are
massive. We can, then, expand them in inverse masses, in which case
all
nonlocal invariants forming the effective action will formally be
expanded in
an infinite series of local invariants
	\begin{eqnarray}
	\int dx\,g^{1/2}\,\sum_{n=0}^{\infty} c_n
	(\underbrace{\nabla...\nabla}_{2n})\,
	\underbrace{R_{\bullet\bullet}R_{\bullet\bullet}...
	R_{\bullet\bullet}}
	_{N}+{\rm O}[R_{\bullet\bullet}^{N+1}].       \label{eqn:3.1}
	\end{eqnarray}
Here the covariant derivatives somehow acting on the product of $N$
Ricci
tensors can be commuted freely because the contribution of their
commutator is
already ${\rm O}[R_{\bullet\bullet}^{N+1}]$. Among
$(\nabla...\nabla)$ only a
limited number $(\leq 2N)$ of derivatives can have indices contracted
with the
indices of
$R_{\bullet\bullet}R_{\bullet\bullet}...R_{\bullet\bullet}$, while
the rest of the derivatives contract with each other to form boxes --
covariant
D'Alambertians acting on separate Ricci tensors or their pairs. This
follows
from a trivial identity
	\begin{eqnarray}
	2\nabla_{i}\nabla_{k}=(\nabla_{i}+\nabla_{k})^2-
	\nabla_{i}^2-\nabla_{k}^2
	\equiv\Box_{i+k}-\Box_{i}-\Box_{k}            \label{eqn:3.2}
	\end{eqnarray}
where $\nabla_{i}$ denotes the covariant derivative acting on the
$i$-th Ricci
tensor in the product $R_{\bullet\bullet}...R_{\bullet\bullet}=
R_{1\bullet\bullet}...R_{i\bullet\bullet}...R_{N\bullet\bullet}$, the
same
notation beeing assumed for covariant boxes $\Box_{i}$ and
$\Box_{i+k}$ acting
on the separate $i$-th Ricci tensor and the separate $ik$-th pair of
those
respectively.

The rearrangement of derivatives in (\ref{eqn:3.1}) according to
(\ref{eqn:3.2}) obviously leads to the following general structure of
$N$-th
order nonlocal invariant
	\begin{eqnarray}
	\int dx\,g^{1/2}\,F(\Box_1,...\Box_N,\Box_{1+2},
	\Box_{1+3},...)
	(\underbrace{\nabla...\nabla}_{\leq 2N})\,
	R_{1\bullet\bullet}R_{2\bullet\bullet}...
	R_{N\bullet\bullet},
\label{eqn:3.3}
	\end{eqnarray}
where $F(\Box_1,...\Box_N,\Box_{1+2},\Box_{1+3},...)$ is an
operatorial
function of boxes accumulating the result of the infinite summation
of
derivatives in (\ref{eqn:3.1}). As above the D'Alambertians $\Box_i$
and
$\Box_{i+k}$ here are acting on the corresponding Ricci tensors or
their pairs
and, similarly to eq.(\ref{eqn:2.3}), have the tensor properties
determined by
the tensor nature of the quantity which they act on. This operatorial
function
is a nonlocal formfactor which serves as a coefficient of the
invariant
	$(\underbrace{\nabla...\nabla}_{\leq 2N})\,
	R_{1\bullet\bullet}R_{2\bullet\bullet}...
	R_{N\bullet\bullet}$
-- the member of the basis in the linear space of nonlocal invariants
of order
$N$ referred to in the Introduction.

In the first three orders these form factors take a simpler form due
to the
integration by parts -- the property which, under the asymptotically
flat
boundary conditions providing the absence of surface terms \cite{II},
can be
written as
	\begin{equation}
	\nabla_1+\nabla_2+...\nabla_N=0.
\label{eqn:3.4}
	\end{equation}
For $N=1$ this relation implies that the first-order formfactor is
always a
local constant, $F(\Box_1)=F(0)={\rm const}$. The second-order form
factor is a
function of one box $\Box_1=\Box_2=\Box,\;F(\Box_1,\Box_2)=f(\Box)$,
and the
third-order formfactor is a function of three boxes acting on three
separate
Ricci curvatures:
	\begin{eqnarray}
	F(\Box_1,\Box_2,\Box_{3})
	(\underbrace{\nabla...\nabla}_{\leq 6})\,
	{R_{1\bullet\bullet}R_{2\bullet\bullet}
	R_{3\bullet\bullet}}.
\label{eqn:3.5}
	\end{eqnarray}
The mixed-box arguments $\Box_{1+2},...$ appear in form factors
beginning with
the fourth order because at this order scattering starts, and these
quantities
are responsible for the arguments of the on-shell scattering
amplitudes. In
particular, at $N=4$ the arguments $\Box_{1+2}$, $\Box_{1+3}$ and
$\Box_{1+4}$
exactly correspond to the Mandelstam variables $s=(p_1+p_2)^2$,
$t=(p_1+p_3)^2$
and $u=(p_1+p_4)^2$ of the formfactor rewritten in the momentum
representation.

Thus, the effective action analytic in spacetime curvature can be
represented
as a series in nonlocal invariants (\ref{eqn:3.3}). It should be
mentioned that
certain caution is needed while working with this expansion. In
particular, the
operatorial arguments $\Box_1,\Box_2,...$ commute with each other
because they
act on different functions, but the mixed-box arguments
$\Box_{1+2},\Box_{1+3},...$ do not commute with each other and with
$\Box_1,\Box_2,...$. Therefore, beginning with the fourth order one
must take
trouble in ordering these arguments in a definite way. Also, while
working with
the accuracy ${\rm O}[R_{\bullet\bullet}^{M+1}]$ it is important to
fix in the
lower-order terms (\ref{eqn:3.3}) with $N<M$ the ordering of
derivatives
$(\underbrace{\nabla...\nabla}_{\leq 2N})$ and the form factor
$F(\Box_1,...)$.

Of course, we don't assume that the form factors are expandable in
any local
series. We took this only for the purpose of building the basis of
invariants.
What we shall really assume about the form factors is their analytic
properties. In quantum field theory we get the form factors in the
Euclidean
spacetime, i.e. for real negative $\Box$'s. We next assume that they
can be
analytically continued to the whole of the complex plane in each of
the
arguments, and that their singularities are only at real non-negative
$\Box$.
This allows us to write and use the spectral representation which in
the
simplest case has the form
	\begin{eqnarray}
	F(\Box_1,\Box_2,...)=\int_{0}^{\infty} dm_1^2\, dm_2^2\,...
	\frac{\rho(m_1^2,m_2^2,...)}
	{(m_1^2-\Box_1)\,(m_2^2-\Box_2)\,...},
\label{eqn:3.6}
	\end{eqnarray}
where all the information about non-local form factors is now
contained in the
spectral densities $\rho(m_1^2,m_2^2,...)$ and the only non-local
structure
that remains is contained in massive Green's functions. Generally
this
representation should be modified by subtraction terms accounting for
a
possible growth of the form factors at large boxes \cite{II,III,IV},
while
their growth at small arguments is severely limited by the
requirement of the
asymptotic flatness of the theory \cite{Armen} and provides the
consistency of
the representation (\ref{eqn:3.6}) at small energies.

The spectral forms (\ref{eqn:3.6}) constitute the basis of the
model-independent approach to quantum gravity theory. They encode all
the
unknown information about the fundumental model in the set of
spectral
densities $\rho(m_1^2,m_2^2,...)$ \cite{CQG} and have a number of
advantages.
In particular, they allow one to order the noncommuting arguments of
the form
factors, provide simple variational rules which, even for a nonlocal
operator
function of one self-commuting argument, would generally be a problem
and make
the equations of the theory manageable, since the only nonlocal
ingredient that
enters the formalism is the Green function. Most important, however,
is that
the spectral forms provide a technical means of imposing correct
boundary
conditions in nonlocal effective equations. As shown in
\cite{Frolov,I} these
equations for expectation values of fields in the standard in-vacuum
can be
obtained by varying the Euclidean effective action with a subsequent
replacement of nonlocal Euclidean form factors by their retarded
Lorentzian
analogues (see also Ref.\onlinecite{Hartle} for a similar recipe in
the first
order of the perturbation theory). This procedure boils down to using
the
retarded massive Green functions in the spectral representation
(\ref{eqn:3.6})
of the form factors in {\it effective equations} for expectation
values.

\section{The basis of gravitational invariants to third order}
\indent
Now we shall implement the procedure of the above type to build
explicitly the
basis of Ricci-tensor invariants to third order. To begin with,
consider the
first order $N=1$ in which case the the invariant
	$(\underbrace{\nabla...\nabla}_{\leq 2})\,
	R_{\bullet\bullet}$,
with derivatives not boiling down to a box, obviously reduces to the
Ricci
scalar. Due to vanishing of total derivative terms in asymptotically
flat
spacetime mentioned above (cf. eq.(\ref{eqn:3.4})) its coefficient is
always a
pure numerical constant, so that the first order invariant is
	\begin{equation}
	{\rm const}\times R, \;\;\;
	R\equiv g^{\mu\nu}R_{\mu\nu}
\label{eqn:4.1}
	\end{equation}
Here "const" is a simplest example of an unknown form factor. From
experiment
we know its value -- it is the inverse Newton constant. Generally,
similar
considerations apply to {\it all} vacuum form factors, the experiment
allowing
us to extract a part of information about them and encode in the
spectral
densiuties discussed above. It should be emphasized that not all
details of the
form factors are equally important. In the example above it is only
important
to know that the Newton constant is positive. The above first-order
invariant
is what Einstein incidentally confined himself with. However, this is
not the
full effective action. To learn more about it we must proceed
further.

In the second order the set of invariants
	$(\underbrace{\nabla...\nabla}_{\leq 4})\,
	R_{1\bullet\bullet}R_{2\bullet\bullet}$
with covariant derivatives not forming boxes simplifies to two
structures
	\begin{equation}
	R_{1\mu\nu}R_2^{\mu\nu}, \,\,\,\,R_1R_2,
\label{eqn:4.2}
	\end{equation}
on account of the contracted form of Bianchi identity
$\nabla^{\mu}R_{\mu\nu}=\frac{1}{2}\nabla_{\nu}R$ and the integration
by parts.
As was mentioned above, the coefficients of these structures in the
efective
action are already the nonlocal formfactors of one argument
$f(\Box_1)=f(\Box_2)$.

In the third order the full set of invariants
	$(\underbrace{\nabla...\nabla}_{\leq 6})\,
	R_{1\bullet\bullet}R_{2\bullet\bullet}
	R_{3\bullet\bullet}$
includes the structures with six, four, two and zero number of
derivatives
whose indices are contracted with the indices of Ricci tensors. Let
us
consequitively build all of them. For this purpose note the following
general
rule: in any structure with derivatives saturating the contraction of
Ricci
curvature indices the covariant derivative acting on one of Ricci
curvatures
can be contracted only with the index of another curvature, for
otherwise the
Bianchi identity generates the invariant with two derivatives
contracted with
one another, which in view of (\ref{eqn:3.2}) should be absorbed into
nonlocal
formfactor and don't enter the tensor invariant itself. The same rule
together
with integration by parts says that the following identification can
be assumed
in such invariants
	\begin{eqnarray}
	R_{1\mu\bullet}\nabla^{\mu}R_{2\bullet\bullet}
	R_{3\bullet\bullet}=
	-R_{1\mu\bullet}R_{2\bullet\bullet}
	\nabla^{\mu}R_{3\bullet\bullet}+(...)
\label{eqn:4.3}
	\end{eqnarray}
modulo the box formfactor terms and higher orders of the curvature
denoted by
(...). From this it immeadiately follows that the only independent
invariant
with six derivatives can be cast in the form
	\begin{eqnarray}
	\nabla_\alpha\nabla_\beta R_1^{\gamma\delta}
        \nabla_\gamma\nabla_\delta R_2^{\mu\nu}
        \nabla_\mu\nabla_\nu R_3^{\alpha\beta}.
\label{eqn:4.4}
	\end{eqnarray}

In the invariant with four derivatives one pair of indices belonging
to three
Ricci curvatures must be contracted, the rest of them beeing
contracted with
indices of derivatives. There are two such contractions:
$R_1^{\mu\nu}R_2^{\alpha\beta}R_3$ and
$R_1^{\alpha\lambda}R_{2\lambda}^{\beta}R_3^{\mu\nu}$ which generate
in view of
(\ref{eqn:4.3}) the following independent invariants with four
derivatives
	\begin{eqnarray}
	\nabla_{\alpha}\nabla_{\beta} R_1^{\mu\nu}
	\nabla_{\mu}\nabla_{\nu}R_2^{\alpha\beta}R_3,\;\;\;\;\;\;
	\nabla_{\mu}R_1^{\alpha\lambda}
	\nabla_{\nu}R_{2\,\lambda}^{\beta}
	\nabla_{\alpha}\nabla_{\beta}R_3^{\mu\nu}
\label{eqn:4.5}
	\end{eqnarray}
(the independent invariants with $\nabla_{\alpha}$ or
$\nabla_{\beta}$ acting
respectively on $R_1^{\alpha\lambda}$ or $R_{2\,\lambda}^{\beta}$
above are
ruled out by integration by parts which reduces them to the first of
the
structures (\ref{eqn:4.5})).

There are four tensors with two uncontracted indices
$R_1^{\alpha\beta}R_2
R_3$, $R_1^{\nu\alpha} R_{2\,\mu\alpha} R_3$,
$R_1^{\mu\nu} R_2^{\alpha\beta} R_{3\,\alpha\beta}$ and $R_1^{\mu\nu}
R_{2\,\beta\mu} R_{3\,\nu}^{\alpha}$ that give rise to the following
invariants
with two derivatives
	\begin{eqnarray}
	&&R_1^{\alpha\beta}\nabla_{\alpha}R_2
	\nabla_{\beta}R_3,\;\;\;\;\;\;
	\nabla^{\mu}R_1^{\nu\alpha}
	\nabla_{\nu}R_{2\,\mu\alpha} R_3,
\nonumber\\
	&&R_1^{\mu\nu} \nabla_{\mu}R_2^{\alpha\beta}
	\nabla_{\nu}R_{3\,\alpha\beta},\;\;\;\;\;\;
	R_1^{\mu\nu} \nabla_{\alpha}R_{2\,\beta\mu}
	\nabla^{\beta}R_{3\,\nu}^{\alpha}.
\label{eqn:4.6}
	\end{eqnarray}

Finally, there are three obvious invariants that contain no
derivatives
	\begin{eqnarray}
	R_1R_2R_3,\;\;\;\;\;\;
	R_{1\,\alpha}^{\mu}R_{2\,\beta}^{\alpha}
	R_{3\,\mu}^{\beta},\;\;\;\;\;\;
	R_1^{\mu\nu}R_{2\,\mu\nu}R_3.
\label{eqn:4.7}
	\end{eqnarray}

It should be emphasized that the full set of cubic invariants follows
from the
above structures (\ref{eqn:4.4}) - (\ref{eqn:4.7}) by all possible
permutations
of three curvature labels 1, 2 and 3. But according to our notations
in
eqs.(\ref{eqn:3.3}) and (\ref{eqn:3.5}) such permutations in
	$(\nabla...\nabla)\,
	{R_{1\bullet\bullet}R_{2\bullet\bullet}
	R_{3\bullet\bullet}}$
can always be replaced by the corresponding permutation of arguments
in the
form factor $F(\Box_1,\Box_2,\Box_{3})$, so that the independent set
of
invariants can be fixed once and for all with the enumeration of
curvatures
chosen above. In what follows we shall assume this rule which allows
one to
reduce the set of independent nonlocal invariants with the proper
account for
the permutation of boxes in their form factors.

\section{The basis of matter invariants to third order}
\indent
The consideration of the above type applies also to matter field
invariants
provided that we know the full set of the field strengths of the
theory. We
shall consider a very general set of field strengths which appears in
a class
of field theories where the Hessian of the action, or the inverse
propagator
for the full collection of fields, can be written as
	\begin{eqnarray}
	g^{\mu\nu}\nabla_{\mu}\nabla_{\nu}\delta^A_B
	+(P^A_B-\frac16 R\delta^A_B).
\label{eqn:5.1}
	\end{eqnarray}
This operator is built of a covariant derivative $\nabla_{\mu}$ and a
potential
term and is applied to an arbitrary set of fields $\varphi^{B}(x)$,
so that $A$
and $B$ stand for any set of discrete indices. The covariant
derivative (with
an arbitrary connection) is characterized by its commutator curvature
	\begin{equation}
	(\nabla_{\mu}\nabla_{\nu}-\nabla_{\nu}\nabla_{\mu})
	\varphi^A={\cal R}^A_{B\,\mu\nu}\varphi^B.
\label{eqn:5.2}
	\end{equation}
The matrix $P^{A}_{B}$ in the potential term of (\ref{eqn:5.1}) is
also
arbitrary (the term $\frac16 R \delta^A_B$ with the Ricci scalar is
added for
convenience), and terms of first order in derivatives are omitted
since they
can always be removed by a redefinition of the covariant derivative.
In what
follows we shall use a hat to indicate the matrix quantities acting
in the
vector space of $\varphi^A$ and also denote the matrix trace
operation in this
space by tr:
	\begin{eqnarray}
	&&\delta^A_B=\hat 1,\;\;\; P^A_B=\hat P,\;\;\;
	{\cal R}^A_{B\,\mu\nu}=\hat{\cal R}_{\mu\nu},\,
	{\rm etc.}\nonumber\\
	&&{\rm tr}\,\hat 1=\delta^A_A,\;\;\;
	{\rm tr}\,\hat P=P^A_A,\;\;\;
	{\rm tr}\,\hat P\,\hat {\cal R}_{\mu\nu}=
	P^A_B\,{\cal R}^B_{A\,\mu\nu},\,{\rm etc.}
\label{eqn:5.3}
	\end{eqnarray}

Thus there are three independent inputs in the operator
(\ref{eqn:5.1}):
$g^{\mu\nu}$, $\nabla_{\mu}$ and $\hat P$ - the metric contracting
the second
derivatives, the connection which defines the covariant derivative
and the
potential matrix. They may be regarded as background fields to which
correspond
their field strengths or curvatures. There is the Riemann curvature
associated
with $g_{\mu\nu}$, the commutator curvature (\ref{eqn:5.2})
associated with
$\nabla_{\mu}$ and the potential $\hat P$ which is its own
"curvature". Since
the Riemann tensor was eliminated above in terms of the Ricci
curvature, the
full set of field strengths characterizing the operator
(\ref{eqn:5.1}), for
which we shall use the collective notation $\Re$, includes
	\begin{equation}
	\Re=(R_{\mu\nu},\;\hat{\cal R}_{\mu\nu},\;\hat P).
\label{eqn:5.4}
	\end{equation}
Purely gravitational invariants to third order in the Ricci tensor
have been
constructed above, so let us consider now the full set of invariants
involving
the matter field strengths $\hat {\cal R}_{\mu\nu}$ and $\hat P$ to
the cubic
order in $\Re$.

In the first order the matter field strengths can contribute only one
invariant
which, for the same reasons as in the gravitational case, can enter
the
effective action of the theory only with a local purely numerical
coefficient:
	\begin{equation}
	{\rm const} \times {\rm tr}\,\hat P
\label{eqn:5.5}
	\end{equation}

In the second order, modulo total derivative, there are only two
invariants of
the form
	$(\underbrace{\nabla...\nabla}_{\leq 4})
	\hat{\cal R}_{1\bullet\bullet}
	\hat{\cal R}_{2\bullet\bullet}$
in which derivatives do not form boxes or commutators leading to
extra power of
$\Re$: ${\rm tr}\,\hat{\cal R}_{1\mu\nu}\hat{\cal R}_2^{\mu\nu}$ and
${\rm tr}\,\nabla^{\mu}\nabla_{\alpha}\hat{\cal R}_{1\mu\nu}\hat{\cal
R}_2^{\alpha\nu}$. However, the cyclic Jacobi identity for the
commutator
curvature
	\begin{equation}
	\nabla_{\alpha}\hat{\cal R}_{\mu\nu}+
	\nabla_{\mu}\hat{\cal R}_{\nu\alpha}+
	\nabla_{\nu}\hat{\cal R}_{\alpha\mu}=0
\label{eqn:5.6}
	\end{equation}
leads to the relation
	\begin{equation}
	\nabla_{\alpha}\hat{\cal R}_{\mu\nu}
	\hat{\cal R}^{\alpha\nu}=
	\frac12 \nabla_{\mu}\hat{\cal R}_{\nu\alpha}
	\hat{\cal R}^{\nu\alpha},
\label{eqn:5.7}
	\end{equation}
which reduces the second invariant up to a box coefficient to the
first one:
	${\rm tr}\,\hat{\cal R}_{1\mu\nu}
	\hat{\cal R}_2^{\mu\nu}$.
The rest of the second-order invariants include
	${\rm tr}\,\hat P_1 \hat P_2,\;
	 {\rm tr}\,\hat P_1 R_2$,
because in view of integration by parts and the contracted Bianchi
identity
they also exhausts the other possible mixed gravity-matter structure
${\rm
tr}\,R_1^{\mu\nu} \nabla_{\mu}\nabla_{\nu}\hat P_2$. Thus, together
with
(\ref{eqn:4.2}), the basis of structures forming all quadratic
invariants
consists of
\begin{eqnarray}
\Re_1\Re_2({1})&=&
R_{1\,\mu\nu} R_2^{\mu\nu}\hat{1},\nonumber\\
\Re_1\Re_2({2})&=&R_1 R_2\hat{1},\nonumber\\
\Re_1\Re_2({3})&=&\hat{P}_1 R_2,\nonumber\\
\Re_1\Re_2({4})&=&\hat{P}_1\hat{P}_2,\nonumber\\
\Re_1\Re_2({5})&=&\hat{\cal R}_{1\mu\nu}
\hat{\cal R}_2^{\mu\nu}.                             \label{eqn:5.8}
\end{eqnarray}

The third order is much richer for it includes all possible cubic
combinations
of $R_{\mu\nu}$, $\hat{\cal R}_{\mu\nu}$ and $\hat P$. Let us first
consider
$\hat P^3$ and $\hat{\cal R}_{\bullet\bullet}^3$ invariants. The
single $\hat
P^3$ invariant is trivial
		${\rm tr}\,\hat P_1\hat P_2\hat P_3$,
while the whole hirearchy of structures
	$(\underbrace{\nabla...\nabla}_{\leq 6})
	\hat{\cal R}_{1\bullet\bullet}
	\hat{\cal R}_{2\bullet\bullet}
	\hat{\cal R}_{3\bullet\bullet}$
includes apriori the invariants with 6, 4, 2 and zero number of
derivatives.
However, the Jacobi identity (\ref{eqn:5.6}) essentially reduces
their number.
Consider first the invariant with six derivatives. Up to total
derivative terms
and commutators of $\nabla$'s it can be represented as
	${\rm tr}\,\nabla_{\mu}\hat J^{\alpha}_{1}
	\nabla_{\alpha}\hat J^{\beta}_{2}
	\nabla_{\beta}\hat J^{\mu}_{3}$
in terms of the transverse vector
	\begin{equation}
	\hat J^{\alpha}=\nabla_{\lambda}
	\hat{\cal R}^{\lambda\alpha},\;\;\;\;
	\nabla_{\alpha}\hat J^{\alpha}=0.      \label{eqn:5.11}
	\end{equation}
In view of (\ref{eqn:5.6}) this vector satisfies the relation
	\begin{eqnarray}
	\nabla_{\mu}\hat J^{\alpha}=
	\nabla^{\alpha}\!\hat J_{\mu}-
	\Box\,\hat{\cal R}^{\alpha}_{.\,\mu}+O\,[\,\Re^2],
\label{eqn:5.12}
	\end{eqnarray}
which means that the above cubic invariant with six derivatives not
forming
boxes is absent. The set of invariants with four derivatives again
reduces to
the only one structure
	${\rm tr}\,\nabla_{\mu}\hat J_{1\nu}
	\hat J^{\mu}_{2} \hat J^{\nu}_{3}$
which does not straightforwardly disappear (up to irrelevant terms)
due to
eqs.(\ref{eqn:5.6}) and (\ref{eqn:5.12}). However, in view of
(\ref{eqn:5.12})
one can rewrite this structure as
	${\rm tr}\,\nabla_{\mu}\hat J_{1\nu}
	\hat J^{(\mu}_{2} \hat J^{\nu)}_{3}$,
integrate it by parts and again use (\ref{eqn:5.12}) applied to $\hat
J_2$ and
$\hat J_3$. Then it takes the form
	$-\frac12\,\hat J_{1\nu}
	\nabla^{\nu}(\hat J^{\mu}_{2} \hat J_{3\mu})$
which disappears up to the total derivative term in view of the
transversality
of (\ref{eqn:5.11}). Thus, there is no invariant with four
derivatives either.
The invariants with two derivatives originate from all possible
differentiations of the following two structures
	${\rm tr}\,\hat{\cal R}_1^{\mu\nu}
	\hat{\cal R}_{2}^{\alpha\beta}
	\hat{\cal R}_{3\alpha\beta}$
and
	${\rm tr}\,\hat{\cal R}_{1}^{\alpha\beta}
	\hat{\cal R}_{2\,\alpha}^{\mu}
	\hat{\cal R}_{3\,\beta}^{\nu}$
and, by integration by parts and use of the Jacobi identity, boil
down to
	${\rm tr}\,\hat{\cal R}_1^{\mu\nu}
	\nabla_{\mu}\hat{\cal R}_{2}^{\alpha\beta}
	\nabla_{\nu}\hat{\cal R}_{3\alpha\beta}$
and
	${\rm tr}\,\hat{\cal R}_{1}^{\alpha\beta}
	\hat J_{2\,\alpha}
	\hat J_{3\,\beta}$.
But the first of these invariants reduces to the second one by the
following
sequence of transformations. First integrate it by parts and use
(\ref{eqn:5.7}) to convert it to the form
	$2\,{\rm tr}\, \hat J_1^{\nu}
	\hat{\cal R}_2^{\alpha\beta}
	\nabla_{\alpha}\hat{\cal R}_{3\beta\nu}$.
Then integrate by parts again and use (\ref{eqn:5.12}) to interchange
the
indices in $\nabla_{\alpha}\hat J_1^{\nu}$. The sequence of a new
integration
by parts and the use of (\ref{eqn:5.7}) eventually leads to the
equation
\mathindent=0pt
\arraycolsep=0pt
	\begin{eqnarray}
	{\rm tr}\,\hat{\cal R}_1^{\mu\nu}
	\nabla_{\mu}\hat{\cal R}_{2}^{\alpha\beta}
	\nabla_{\nu}\hat{\cal R}_{3\alpha\beta}&=&
	2\,{\rm tr}\,\hat{\cal R}_{3}^{\alpha\beta}
	\hat J_{1\,\alpha}\hat J_{2\,\beta}
	+2\,{\rm tr}\,\hat{\cal R}_{2}^{\alpha\beta}
	\hat J_{3\,\alpha}\hat J_{1\,\beta}
	+2\,{\rm tr}\,\Box\,\hat{\cal R}_{1\,.\beta}^{\alpha}
	\hat{\cal R}_{2\,.\nu}^{\beta}
	\hat{\cal R}_{3\,.\alpha}^{\nu}   \nonumber\\
	&&\qquad\quad
	-{\rm tr}\,\hat{\cal R}_1^{\mu\nu}
	\nabla_{\mu}\hat{\cal R}_{2}^{\alpha\beta}
	\nabla_{\nu}\hat{\cal R}_{3\alpha\beta}
	+O\,[\,\Re^4]+{\rm a\;total\;derivative},
\label{eqn:5.13}
	\end{eqnarray}
\arraycolsep=3pt
\mathindent=\leftmargini
which can be easily solved for
	${\rm tr}\,\hat{\cal R}_1^{\mu\nu}
	\nabla_{\mu}\hat{\cal R}_{2}^{\alpha\beta}
	\nabla_{\nu}\hat{\cal R}_{3\alpha\beta}$
in terms of the invariant
	${\rm tr}\,\hat{\cal R}_{1}^{\alpha\beta}
	\nabla^{\mu}\hat {\cal R}_{2\,\mu\alpha}
	\nabla^{\nu}\hat {\cal R}_{3\,\nu\beta}$
and the only possible $\hat{\cal R}^3$-invariant without derivatives
	${\rm tr}\,\hat{\cal R}_{1.\,\beta}^{\;\alpha}
	\hat{\cal R}_{2.\,\nu}^{\;\beta}
	\hat{\cal R}_{3.\,\alpha}^{\;\nu}$
(under a proper permutation of curvature labels, cf. discussion at
the end of
Sect.IV).

A similar technique can be applied for a construction of all the rest
gravity-matter invariants. Again their number is essentially reduced
by using
the identities of the above type. Here are some examples of such a
reduction \cite{IV}
which we present without derivation starting with a corollary of
(\ref{eqn:5.13})
\mathindent=0pt
\arraycolsep=0pt
	\begin{eqnarray}
	&&{\rm tr}\,
  	\hat{\cal R}_1^{\alpha\beta}
  	\nabla_\alpha\hat{\cal R}_2^{\mu\nu}
  	\nabla_\beta\hat{\cal R}_{3\mu\nu}={\rm tr}\Big(
	\Box_1
\hat{\cal R}^\mu_{1\alpha}\hat{\cal R}^\alpha_{2\beta}
\hat{\cal R}^\beta_{3\mu}
\nonumber\\
&&\qquad
+\hat{\cal R}_{3\alpha\beta}
 \nabla_\mu\hat{\cal R}_1^{\mu\alpha}\nabla_\nu
\hat{\cal R}_2^{\nu\beta}
+\hat{\cal R}_{2\alpha\beta}
 \nabla_\mu\hat{\cal R}_3^{\mu\alpha}\nabla_\nu
\hat{\cal R}_1^{\nu\beta}\Big)
+{\rm O}[\Re^4] +\
{\rm a\ total\ derivative},                    \label{eqn:5.16}
\\
&&{\rm tr}\nabla_\mu\nabla_\lambda\hat{\cal R}_1^{\lambda\nu}
\nabla^\alpha\hat{\cal R}_{2\alpha\nu}
\nabla_\sigma\hat{\cal R}_3^{\sigma\mu}=-\frac12{\rm tr}\Big(
\Box_1\hat{\cal R}_{1\alpha\beta}
\nabla_\mu\hat{\cal R}_2^{\mu\alpha}
\nabla_\nu\hat{\cal R}_3^{\nu\beta}
\nonumber\\
&&\qquad
+\Box_2\hat{\cal R}_{2\alpha\beta}
\nabla_\mu\hat{\cal R}_3^{\mu\alpha}
\nabla_\nu\hat{\cal R}_1^{\nu\beta}\!
-\Box_3\hat{\cal R}_{3\alpha\beta}
\nabla_\mu\hat{\cal R}_1^{\mu\alpha}
\nabla_\nu\hat{\cal R}_2^{\nu\beta} \Big)
+{\rm O}[\Re^4] +
{\rm a\ total\ derivative},                    \label{eqn:5.17}
\\
&&{\rm tr}\,\nabla_\alpha R_1^{\beta\mu}
\nabla_\beta\hat{\cal R}_2^{\alpha\nu}
\hat{\cal R}_{3\mu\nu}=
-\frac12 {\rm tr}\,R_1^{\alpha\beta}
\nabla_\alpha\hat{\cal R}_2^{\mu\nu}
\nabla_\beta\hat{\cal R}_{3\mu\nu}
\nonumber\\
&&\qquad
+{\rm tr}\,R_1^{\mu\nu}
\nabla_\mu\nabla_\lambda\hat{\cal R}_2^{\lambda\alpha}
\hat{\cal R}_{3\alpha\nu}
+{\rm O}[\Re^4] +\
{\rm a\ total\ derivative},                    \label{eqn:5.18}
\\
&&{\rm tr}\nabla^\mu R_{1\beta\alpha}
\nabla^\beta\hat{\cal R}_2^{\alpha\nu}
\hat{\cal R}_{3\mu\nu}
={\rm tr}\Big[-\frac18(\Box_1+\Box_2+\Box_3)
R_1\hat{\cal R}_2^{\mu\nu}\hat{\cal R}_{3\mu\nu}
+\frac12 R_1^{\alpha\beta}
\nabla_\alpha\hat{\cal R}_2^{\mu\nu}
\nabla_\beta\hat{\cal R}_{3\mu\nu}
\nonumber\\
&&\qquad
-\frac12R_1\nabla_\alpha\hat{\cal R}_2^{\alpha\mu}
 \nabla^\beta\hat{\cal R}_{3\beta\mu}
-R_1^{\mu\nu}\nabla_\mu\nabla_\lambda
  \hat{\cal R}_3^{\lambda\alpha}\hat{\cal R}_{2\alpha\nu}\Big]
+{\rm O}[\Re^4] +\
{\rm a\ total\ derivative},                      \label{eqn:5.19}
\\
&&{\rm tr} R_1^{\mu\nu}
\nabla_\mu\nabla_\lambda\hat{\cal R}_2^{\lambda\alpha}
\nabla_\nu\nabla^\sigma\hat{\cal R}_{3\sigma\alpha}
={\rm tr}\Big[-\Box_2\Box_3 R_1^{\mu\nu}
  \hat{\cal R}_{2\mu}
{}^\alpha\hat{\cal R}_{3\nu\alpha}
  -\Box_3R_1^{\mu\nu}
 \nabla_\mu\nabla_\lambda\hat{\cal R}_2^{\lambda\alpha}
 \hat{\cal R}_{3\alpha\nu}
\nonumber\\
&&\qquad
+\frac12(\Box_1-\Box_2-\Box_3)R_1^{\mu\nu}
  \nabla^\alpha\hat{\cal R}_{2\alpha\mu}
  \nabla^\beta\hat{\cal R}_{3\beta\nu}
-\Box_2R_1^{\mu\nu}
 \nabla_\mu\nabla_\lambda\hat{\cal R}_3^{\lambda\alpha}
 \hat{\cal R}_{2\alpha\nu}\Big]
\nonumber\\
&&\qquad
+{\rm O}[\,\Re^4]+{\rm a\ total\ derivative}         \label{eqn:5.20}
\end{eqnarray}
\arraycolsep=3pt
\mathindent=\leftmargini

As a result of such a reduction we have for the full basis of
gravitational and
matter invariants eleven structures without derivatives
\begin{eqnarray}
\Re_1\Re_2\Re_3({1})&=&\hat{P}_1\hat{P}_2\hat{P}_3,\nonumber\\
\Re_1\Re_2\Re_3({2})&=&\hat{\cal R}^{\ \mu}_{1\ \alpha}
\hat{\cal R}^{\ \alpha}_{2\ \beta}
\hat{\cal R}^{\ \beta}_{3\ \mu},\nonumber\\
\Re_1\Re_2\Re_3({3})&=&\hat{\cal R}^{\mu\nu}_1
\hat{\cal R}_{2\,\mu\nu}\hat{P}_3,\nonumber\\
\Re_1\Re_2\Re_3({4})&=&R_1 R_2 \hat{P}_3,\nonumber\\
\Re_1\Re_2\Re_3({5})&=&R_1^{\mu\nu}
R_{2\,\mu\nu}\hat{P}_3,\nonumber\\
\Re_1\Re_2\Re_3({6})&=&\hat{P}_1\hat{P}_2 R_3,\nonumber\\
\Re_1\Re_2\Re_3({7})&=&R_1\hat{\cal R}^{\mu\nu}_2
\hat{\cal R}_{3\,\mu\nu},\nonumber\\
\Re_1\Re_2\Re_3({8})&=&R_1^{\alpha\beta}
\hat{\cal R}_{2\,\alpha}^{\ \ \ \,\mu}
\hat{\cal R}_{3\,\beta\mu},\nonumber\\
\Re_1\Re_2\Re_3({9})&=&R_1 R_2 R_3\hat{1},\nonumber\\
\Re_1\Re_2\Re_3({10})&=&R_{1\,\alpha}^\mu
R_{2\,\beta}^{\alpha} R_{3\,\mu}^\beta\hat{1},\nonumber\\
\Re_1\Re_2\Re_3({11})&=&
R_1^{\mu\nu}R_{2\,\mu\nu}R_3\hat{1},               \label{eqn:5.21}
\end{eqnarray}
fourteen structures with two derivatives
\begin{eqnarray}
\Re_1\Re_2\Re_3({12})&=&
\hat{\cal R}_1^{\alpha\beta}\nabla^\mu
\hat{\cal R}_{2\mu\alpha}\nabla^\nu
\hat{\cal R}_{3\nu\beta},\nonumber\\
\Re_1\Re_2\Re_3({13})&=&\hat{\cal R}_1^{\mu\nu}
\nabla_\mu\hat{P}_2\nabla_\nu\hat{P}_3,\nonumber\\
\Re_1\Re_2\Re_3({14})&=&\nabla_\mu
\hat{\cal R}_1^{\mu\alpha}\nabla^\nu
\hat{\cal R}_{2\,\nu\alpha}\hat{P}_3,\nonumber\\
\Re_1\Re_2\Re_3({15})&=&R_1^{\mu\nu}\nabla_\mu R_2\nabla_\nu
\hat{P}_3,\nonumber\\
\Re_1\Re_2\Re_3({16})&=&\nabla^\mu R_1^{\nu\alpha}\nabla_\nu
R_{2\,\mu\alpha}\hat{P}_3,\nonumber\\
\Re_1\Re_2\Re_3({17})&=&R_1^{\mu\nu}\nabla_\mu\nabla_\nu
\hat{P}_2\hat{P}_3,\nonumber\\
\Re_1\Re_2\Re_3({18})&=&R_{1\,\alpha\beta}\nabla_\mu
\hat{\cal R}_2^{\mu\alpha}\nabla_\nu
\hat{\cal R}_3^{\nu\beta},\nonumber\\
\Re_1\Re_2\Re_3({19})&=&R_1^{\alpha\beta}
\nabla_\alpha\hat{\cal R}_2^{\mu\nu}\nabla_\beta
\hat{\cal R}_{3\,\mu\nu},\nonumber\\
\Re_1\Re_2\Re_3({20})&=&R_1\nabla_\alpha
\hat{\cal R}_2^{\alpha\mu}\nabla^\beta
\hat{\cal R}_{3\,\beta\mu}\nonumber,\\
\Re_1\Re_2\Re_3({21})&=&R_1^{\mu\nu}\nabla_\mu
\nabla_\lambda\hat{\cal R}_2^{\lambda\alpha}
\hat{\cal R}_{3\,\alpha\nu},\nonumber\\
\Re_1\Re_2\Re_3({22})&=&R_1^{\alpha\beta}
\nabla_\alpha R_2 \nabla_\beta R_3\hat{1},\nonumber\\
\Re_1\Re_2\Re_3({23})&=&\nabla^\mu R_1^{\nu\alpha}
\nabla_\nu R_{2\,\mu\alpha}R_3\hat{1},\nonumber\\
\Re_1\Re_2\Re_3({24})&=&R_1^{\mu\nu}
\nabla_\mu R_2^{\alpha\beta}\nabla_\nu R_{3\,\alpha\beta}
\hat{1},\nonumber\\
\Re_1\Re_2\Re_3({25})&=&R_1^{\mu\nu}
\nabla_\alpha R_{2\,\beta\mu}\nabla^\beta
R_{3\,\nu}^\alpha\hat{1},                          \label{eqn:5.22}
\end{eqnarray}
three structures with four derivatives
\begin{eqnarray}
\Re_1\Re_2\Re_3({26})&=&\nabla_\alpha\nabla_\beta
R_1^{\mu\nu}\nabla_\mu\nabla_\nu
R_2^{\alpha\beta}\hat{P}_3,\nonumber\\
\Re_1\Re_2\Re_3({27})&=&\nabla_\alpha\nabla_\beta
R_1^{\mu\nu}\nabla_\mu\nabla_\nu R_2^{\alpha\beta}
R_3\hat{1},\nonumber\\
\Re_1\Re_2\Re_3({28})&=&\nabla_\mu
R_1^{\alpha\lambda} \nabla_\nu
R_{2\,\lambda}^\beta\nabla_\alpha\nabla_\beta
R_3^{\mu\nu}\hat{1},                               \label{eqn:5.23}
\end{eqnarray}
one structure with six derivatives
\begin{eqnarray}
\Re_1\Re_2\Re_3({29})&=&\nabla_\lambda\nabla_\sigma
R_1^{\alpha\beta}\nabla_\alpha\nabla_\beta
R_2^{\mu\nu}
\nabla_\mu\nabla_\nu R_3^{\lambda\sigma}\hat{1},     \label{eqn:5.24}
\end{eqnarray}
and, finally, there is a set of additional nine structures linear in
$\hat
{\cal R}_{\mu\nu}$
	\begin{eqnarray}
	\Re_1\Re_2\Re_3({30})&=&R_{1\,\alpha}^\mu
	R_{2\,\beta}^{\alpha}
	\hat {\cal R}_{3\,\mu}^\beta,\nonumber\\
	\Re_1\Re_2\Re_3({31})&=&\hat{P}_1 \nabla_\mu \hat{\cal
	R}_{2\,\nu\alpha}
	\nabla^\nu R_3^{\mu\alpha},\nonumber\\
	\Re_1\Re_2\Re_3({32})&=&\hat{P}_1 \nabla_\beta \hat{\cal
	R}_2^{\beta\alpha}
	\nabla_\alpha R_3,\nonumber\\
	\Re_1\Re_2\Re_3({33})&=&
	\nabla_\mu \hat{\cal R}_1^{\mu\alpha}
	R_{2\,\alpha\beta}\nabla^\beta\hat{P}_3,\nonumber\\
	\Re_1\Re_2\Re_3({34})&=&\nabla_\mu \hat{\cal R}_1^{\mu\alpha}
	R_{2\,\alpha\beta}\nabla^\beta R_3,\nonumber\\
	\Re_1\Re_2\Re_3({35})&=&R_1 \nabla_\mu \hat{\cal
	R}_{2\,\nu\alpha}
	\nabla^\nu R_3^{\mu\alpha},\nonumber\\
	\Re_1\Re_2\Re_3({36})&=&\nabla_\beta
	\hat{\cal R}_1^{\beta\alpha}
	\nabla_\alpha R_2 R_3,\nonumber\\
	\Re_1\Re_2\Re_3({37})&=&\hat{\cal R}_1^{\alpha\beta}
	\nabla_\mu R_{2\,\nu\alpha}
	\nabla^\nu R_{3\,\beta}^{\mu},\nonumber\\
	\Re_1\Re_2\Re_3({38})&=&\nabla^\alpha\hat{\cal
R}_1^{\beta\lambda}
	\nabla_\lambda\nabla_\mu R_{2\,\nu\alpha}
	\nabla^\nu R_{3\,\beta}^{\mu}.
\label{eqn:5.25}
	\end{eqnarray}

Thus, the structures (\ref{eqn:4.1}), (\ref{eqn:5.5}),
(\ref{eqn:5.8}) and
(\ref{eqn:5.21}) - (\ref{eqn:5.25}) form a complete basis of nonlocal
invariants to third order in the curvature. The background-field
functionals
belonging to the class of invariants discussed in Introduction have,
in the
notations of Sect.III, the expansion of the form
	\begin{eqnarray}
	&&\int dx\, g^{1/2}\, \,{\rm tr}\,
	\Big\{\,{\rm const}\,R\hat{1}+{\rm const}\,\hat{P}
	+\sum^{5}_{i=1} f_{i\;}(\Box_2)\,
	\Re_1\Re_2({i})\nonumber\\
	&&\qquad\qquad\qquad\qquad\qquad\qquad
	+\sum^{38}_{i=1} F_{i}(\Box_1,\Box_2,\Box_3)\,
	\Re_1\Re_2\Re_3({i})+{\rm O}[\,\Re^4]\,\Big\}.
\label{eqn:5.26}
	\end{eqnarray}

In the third order, which is a matter of our prime interest here,
there are
thirty eight invariants admissible by the requirements of covariance
and
asymptotic flatness. (Ten of them are purely gravitational, and with
gravity
switched off there are six.) In the field-theoretic calculations,
however, the
last nine structures have a special status which makes us to present
them in a
separate list (\ref{eqn:5.25}). As shown in \cite{IV}, the trace of
the heat
kernel for the operator (\ref{eqn:5.1}) and the corresponding
one-loop
effective action do not contain these structures because their form
factors
$F_{i}(\Box_1,\Box_2,\Box_3),\,i=30,...38,$ either identically vanish
or have
such symmetries under the permutation of the box arguments that make
their
contribution vanishing in view of the symmetries of the structures
themselves.
For example, the structure 36 is antisymmetric in labels 2 and 3,
	$\nabla_\beta
	\hat{\cal R}_1^{\beta\alpha}\nabla_\alpha R_2 R_3
	=-\nabla_\beta \hat{\cal R}_1^{\beta\alpha}
	\nabla_\alpha R_3 R_2+{\rm a\ total\ derivative}$,
while its heat-kernel form factor is symmetric \cite{IV},
$F_{36}(\Box_1,\Box_2,\Box_3)=F_{36}(\Box_1,\Box_3,\Box_2)$. The same
symmetry
of the heat-kernel form factor $F_{13}(\Box_1,\Box_2,\Box_3)$ does
not,
however, annihilate the contribution of the structure
$\Re_1\Re_2\Re_3({13})$
linear in $\hat {\cal R}_{\mu\nu}$, because all of its three
curvatures are
matrices, and its antisymmetrization in the labels 2 and 3 is
proportional to
the nonvanishing commutator $\Big[\nabla_\mu \hat{P},\nabla_\nu
\hat{P}\Big]$
(see Refs. \onlinecite{IV,freaks}).

\section{Hidden identities between nonlocal cubic invariants}
\indent
For low-dimensional manifolds the basis built above may be redundant.
Thus, in
the two-dimensional case $2\omega=2$, because of the identity
$R_{\mu\nu}=\frac12\,g_{\mu\nu}R$, there is just one purely
gravitational
nonlocal structure at each order
	\begin{eqnarray}
	\int dx\,g^{1/2}\,F(\Box_1,...\Box_N,\Box_{1+2},
	\Box_{1+3},...)
	\,R_{1}R_{2}...R_{N},
\label{eqn:6.1}
	\end{eqnarray}
and, moreover, at first order there is none because the Ricci scalar
is a total
derivative. For $2\omega\geq 2$ the first- and second-order bases are
already
irreducible but the third-order basis is not. To show this, we begin
the
analysis of cubic invariants with the local identities for a tensor
possessing
the symmetries of the Weyl tensor:
\begin{eqnarray}
&&C_{\alpha\beta\gamma\delta} =
C_{[\alpha\beta]\gamma\delta} =
C_{\alpha\beta[\gamma\delta]} ,\;\;\;\;\;
C_{\alpha\beta\gamma\delta} =
C_{\gamma\delta\alpha\beta} ,              \label{eqn:6.2}\\
&&C_{\alpha\beta\gamma\delta} +
C_{\alpha\delta\beta\gamma} +
C_{\alpha\gamma\delta\beta} = 0 ,\;\;
C^\alpha{}_{\beta\alpha\delta} = 0.         \label{eqn:6.3}
\end{eqnarray}
Here the Ricci identity (\ref{eqn:6.3}) can be rewritten as
	\begin{equation}
	C_{\alpha[\beta\gamma]\delta} =
	-\frac12C_{\alpha\delta\beta\gamma}   \label{eqn:6.4}
	\end{equation}
and, with the use of (\ref{eqn:6.2}), as
	\begin{equation}
	C_{\alpha[\beta\gamma\delta]} = 0      \label{eqn:6.5}
	\end{equation}
where the complete antisymmetrization in three
indices is meant. Eq. (\ref{eqn:6.4}) is useful when forming
contractions because it shows that the contractions
of the form
$C^{\,\cdot\,\alpha\beta.}
C_{\,\cdot\,\cdot\,\alpha\beta} {\rm\ and\ }
C^{\,\cdot\,\cdot\,\alpha\beta}C_{\,\cdot\,\cdot\,\alpha\beta}$
express through one another.

In view of applications to the gravitational equations
\cite{Fr-Zeln}, we first list all possible cubic contractions
having two free indices. The symmetries above allow only
\begin{eqnarray}
J_{1}{}^\nu_\mu&=&C_{\mu\beta\gamma\delta}
       C^{\nu\beta\alpha\sigma}
       C_{\alpha\sigma}{}^{\gamma\delta},    \label{eqn:6.6}\\
J_{2}{}^\nu_\mu&=&C_{\mu\beta\gamma\delta}
       C^{\nu\alpha\gamma\sigma}
       C_\alpha{}^\beta{}_\sigma{}^\delta,    \label{eqn:6.7}\\
J_{3}{}^\nu_\mu&=&C_{\mu\beta\gamma\delta}
       C^{\nu\alpha\gamma\sigma}
       C_\sigma{}^\beta{}_\alpha{}^\delta,    \label{eqn:6.8}\\
J_{4}{}^\nu_\mu&=&C_{\mu\gamma\beta\delta}
       C^{\nu\beta\alpha\sigma}
       C_{\alpha\sigma}{}^{\gamma\delta},      \label{eqn:6.9}\\
J_{5}{}^\nu_\mu&=&C^\nu{}_{\sigma\mu\kappa}
       C^{\sigma\alpha\beta\gamma}
       C^\kappa{}_{\alpha\beta\gamma},         \label{eqn:6.9a}
\end{eqnarray}
and, furthermore, by (\ref{eqn:6.4}), one has
	\begin{equation}
	J_{4}{}^\nu_\mu=\frac12J_{1}{}^\nu_\mu,  \label{eqn:6.9b}
	\end{equation}
and, by applying the Ricci identity to the last $C$
in (\ref{eqn:6.7}), one obtains
	\begin{equation}
	J_{3}{}^\nu_\mu=J_{2}{}^\nu_\mu
	-\frac12J_{4}{}^\nu_\mu=
	J_{2}{}^\nu_\mu-\frac14J_{1}{}^\nu_\mu.  \label{eqn:6.10}
	\end{equation}
Thus, for an arbitrary space-time dimension $2\omega$,
there are three independent contractions:
$J_{1}{}^\nu_\mu, J_{2}{}^\nu_\mu, J_{5}{}^\nu_\mu$.

For a particular space-time dimension, the number of
independent contractions can be smaller because of the
existence of identities obtained by antisymmetrization
of $(2\omega+1)$ indices. Note that such an antisymmetrization
must not involve more than two indices of each $C$ tensor;
otherwise the identity will be satisfied trivially by
virtue of (\ref{eqn:6.5}). Hence, for three $C$ tensors, the number
of indices involved in the antisymmetrization should not
exceed six, and, therefore, the space-time dimension
$2\omega$ for which nontrivial identities exist cannot
exceed five. For $2\omega\leq5$ we have
\begin{equation}
C_{[\alpha\beta}{}^{\gamma\delta}
C_{\gamma\delta}{}^{\kappa\mu}
C_{\kappa\nu]}{}^{\alpha\beta}
\equiv0,\ \ \ \  2\omega\leq5                   \label{eqn:6.11}
\end{equation}
with the complete antisymmetrization of six lower
indices. When written down explicitly, this
identity takes the form
\begin{equation}
J_{1}{}^\nu_\mu-4J_{3}{}^\nu_\mu
-2J_{5}{}^\nu_\mu=0,\ \ \ \  2\omega\leq5       \label{eqn:6.12}
\end{equation}
or, by (\ref{eqn:6.10}), the form
\begin{equation}
J_{2}{}^\nu_\mu=\frac12J_{1}{}^\nu_\mu
-\frac12J_{5}{}^\nu_\mu,
\ \ \ \  2\omega\leq5                            \label{eqn:6.13}
\end{equation}
and reduces the number of independent contractions down
to two: $J_{1}{}^\nu_\mu$ and $J_{5}{}^\nu_\mu$.

Finally, for $2\omega=4$ (the lowest dimension in which
a nonvanishing Weyl tensor exists), the identity (\ref{eqn:6.11})
becomes a linear combination of the identities
\begin{equation}
C_{[\alpha\beta}{}^{\gamma\delta}
C_{\gamma\delta}{}^{\kappa\mu}
C_{\kappa]\nu}{}^{\alpha\beta}
\equiv0,\ \ \ \  2\omega=4                        \label{eqn:6.14}
\end{equation}
with the antisymmetrization over only five indices,
and there is one more identity, quadratic in $C$ :
\begin{equation}
C_{[\alpha\beta}{}^{\gamma\delta}
C_{\gamma\delta}{}^{\alpha\beta}
\delta_{\mu]}^\nu
\equiv0,\ \ \ \  2\omega=4.                       \label{eqn:6.16}
\end{equation}
Its explicit form is
\begin{equation}
C^{\alpha\beta\gamma\nu}
C_{\alpha\beta\gamma\mu}
=\frac14\delta^\nu_\mu
C_{\alpha\beta\gamma\delta}
C^{\alpha\beta\gamma\delta}
,\ \ \ \  2\omega=4.                             \label{eqn:6.17}
\end{equation}
When this relation is used in (\ref{eqn:6.9a}), the result is
	\begin{equation}
	J_{5}{}^\nu_\mu=0,\ \ \ \  2\omega=4      \label{eqn:6.18}
	\end{equation}
by the second of equations (\ref{eqn:6.3}). Thus, in four dimensions,
there is
only one independent contraction: $J_{1}{}^\nu_\mu$.

Similar results hold for invariants except that the
complete contraction of $J_5$ in (\ref{eqn:6.9a}) is zero for
any space-time dimension. Therefore, initially one has
four different $C^3$ invariants
\begin{eqnarray}
I_1&=&C_{\mu\beta\gamma\delta}
       C^{\mu\beta\alpha\sigma}
       C_{\alpha\sigma}{}^{\gamma\delta},        \label{eqn:6.19}\\
I_2&=&C_{\mu\beta\gamma\delta}
       C^{\mu\alpha\gamma\sigma}
       C_\alpha{}^\beta{}_\sigma{}^\delta,       \label{eqn:6.20}\\
I_3&=&C_{\mu\beta\gamma\delta}
       C^{\mu\alpha\gamma\sigma}
       C_\sigma{}^\beta{}_\alpha{}^\delta,       \label{eqn:6.21}\\
I_4&=&C_{\mu\gamma\beta\delta}
       C^{\mu\beta\alpha\sigma}
       C_{\alpha\sigma}{}^{\gamma\delta}          \label{eqn:6.22}
\end{eqnarray}
with the relations
\begin{equation} I_3=I_2-\frac14I_1,\
I_4=\frac12I_1,                                \label{eqn:6.23}
\end{equation}
and, for $2\omega\leq5$, the identity (\ref{eqn:6.11})
(contracted in $\mu,\nu$) adds one more relation:
	\begin{equation}
	I_2=\frac12I_1,\ \ \ \  2\omega\leq5.    \label{eqn:6.23a}
	\end{equation}
When going over from $2\omega=5$ to $2\omega=4$,
the identity (\ref{eqn:6.17}) leads to no further reduction.
Thus, the dimension of the basis of local $C^3$
invariants is 2 for $2\omega>5$, and 1 for both
$2\omega=5$ and $2\omega=4$.

For invariants with the Riemann tensor, the counting
is different because, in this case, the quadratic
identity (\ref{eqn:6.17}) begins working. For $2\omega\leq5$,
the identity (\ref{eqn:6.23a}) with the Weyl tensor expressed
through the Riemann tensor reduces the number of
independent cubic invariants by one. For $2\omega=4$,
the identity (\ref{eqn:6.17}) contracted with the Ricci tensor:
\begin{equation}
C^{\alpha\beta\gamma\nu}
C_{\alpha\beta\gamma\mu}
R^\mu_\nu
=\frac14R
C_{\alpha\beta\gamma\delta}
C^{\alpha\beta\gamma\delta}
,\ \ \ \  2\omega=4                       \label{eqn:6.24}
\end{equation}
reduces this number by one more. These results
agree with the group-theoretic analysis carried
out in \cite{26} (for the application of this analysis in a geometric
classification of conformal anomalies see Ref.\onlinecite{DesSch}).
According
to \cite{26}, the dimension
of the basis of local cubic invariants with the
Riemann tensor (without derivatives) is 8 for
$2\omega>5$, 7 for $2\omega=5$, and 6 for $2\omega=4$.

We shall now concentrate on the space-time
dimension $2\omega=4$ and go over to the consideration
of {\it nonlocal} invariants cubic in the curvature.
Owing to their algebraic nature, the identities above
admit easily a nonlocal generalization. Indeed, the
two cubic identities obtained by antisymmetrizations in four
dimensions: eq. (\ref{eqn:6.14}) contracted in $\mu,\nu$, and
eq. (\ref{eqn:6.16}) contracted with $R^\mu_\nu$ can
in fact be
written down for three {\em different} tensors
and, in particular, for the curvature tensors at
{\em three different points}. One can then multiply
them by arbitrary form factors and next make the
points coincident. Since the basis of nonlocal invariants above was
built in
terms of Ricci curvatures, rather than the Weyl tensors, we shall
start such a
derivation with equations (\ref{eqn:6.14}) - (\ref{eqn:6.16})
containing
instead of the Weyl tensors the Riemann ones (which will be later
eliminated in
terms of Ricci curvatures). The nonlocal identities obtained in this
way are of
the form
\begin{eqnarray}
&&\widetilde{\cal F}(\Box_1,\Box_2,\Box_3)
R_{1[\alpha\beta}{}^{\gamma\delta}
R_{2\gamma\delta}{}^{\kappa\mu}
R_{3\kappa]\mu}{}^{\alpha\beta}
=0,\ \ \ 2\omega=4,                    \label{eqn:6.25a}\\
&&\widetilde{\widetilde{\cal F}}
(\Box_1,\Box_2,\Box_3)
R_{1[\alpha\beta}{}^{\gamma\delta}
R_{2\gamma\delta}{}^{\alpha\beta}
R_{3\mu]}^\mu=0,\ \ \ \ \ \ \ 2\omega=4,         \label{eqn:6.25}
\end{eqnarray}
with arbitrary $\widetilde{\cal F}(\Box_1,\Box_2,\Box_3)$
and $\widetilde{\widetilde{\cal F}}(\Box_1,\Box_2,\Box_3)$.
Since nothing is involved here except inexistence
of five different indices in four dimensions, these
identities are obviously correct in the present
nonlocal form as well.

When written down explicitly, the left-hand sides of eqs.
(\ref{eqn:6.25a}),
(\ref{eqn:6.25}) take the form (for arbitrary dimension)
\mathindent=0pt
\arraycolsep=0pt
\begin{eqnarray}&&
\widetilde{\cal F}(\Box_1,\Box_2,\Box_3)
R_{1[\alpha\beta}{}^{\gamma\delta}
R_{2\gamma\delta}{}^{\kappa\mu}
R_{3\kappa]\mu}{}^{\alpha\beta}\equiv
\frac1{15}\widetilde{\cal F}(\Box_1,\Box_2,\Box_3)
\Big[ R_{1\alpha\beta\gamma\delta}
R_2^{\gamma\delta\mu\nu}R_{3\mu\nu}{}^{\alpha\beta}
\nonumber\\
&&\ \ \ \ \ \ \ \
-2R_{1\alpha\beta\gamma\delta}
  R_2^{\alpha\mu\gamma\nu}R_3{}^\beta{}_\mu{}^\delta{}_\nu
-3R_{1\alpha\beta}R_2^{\alpha\pi\rho\kappa}
  R_3^\beta{}_{\pi\rho\kappa}
-R_{2\alpha\beta}R_1^{\alpha\pi\rho\kappa}
  R_3^\beta{}_{\pi\rho\kappa}
-R_{3\alpha\beta}R_1^{\alpha\pi\rho\kappa}
  R_2^\beta{}_{\pi\rho\kappa}
\nonumber\\
&&\ \ \ \ \ \ \ \
+\frac12
 R_1R_{2\alpha\beta\gamma\delta}
    R_3^{\alpha\beta\gamma\delta}
+2R_{2\alpha\beta\gamma\delta}
  R_1^{\alpha\gamma}R_3^{\beta\delta}
+2R_{3\alpha\beta\gamma\delta}
  R_1^{\alpha\gamma}R_2^{\beta\delta}
\nonumber\\
&&\ \ \ \ \ \ \ \
+2R_{1\alpha\beta}R_2^{\beta\gamma}R_3{}^\alpha_\gamma
- R_1R_{2\alpha\beta}R_3^{\alpha\beta}
\Big],                                  \label{eqn:6.26a}\\
&&\widetilde{\widetilde{\cal F}}
(\Box_1,\Box_2,\Box_3)
R_{1[\alpha\beta}{}^{\gamma\delta}
R_{2\gamma\delta}{}^{\alpha\beta}
R_{3\mu]}^\mu\equiv
-\frac2{15}\widetilde{\widetilde{\cal F}}
(\Box_1,\Box_2,\Box_3)\,
\Big[
R_{1\alpha\beta\gamma\mu}
R_2^{\alpha\beta\gamma}{}_\nu R_3^{\mu\nu}
\nonumber\\
&&\ \ \ \ \ \ \ \ -\frac14R_{1\alpha\beta\gamma\delta}
    R_2^{\alpha\beta\gamma\delta}R_3
-2R_{1\alpha\beta}R_2^{\beta\gamma}R_3{}^\alpha_\gamma
-R_{1\alpha\beta\gamma\delta}
  R_2^{\alpha\gamma}R_3^{\beta\delta}
-R_{2\alpha\beta\gamma\delta}
  R_1^{\alpha\gamma}R_3^{\beta\delta}
\nonumber\\&&\ \ \ \ \ \ \ \
+R_{1\alpha\beta}R_2^{\alpha\beta}R_3
+\frac12R_1R_{2\alpha\beta}R_3^{\alpha\beta}
+\frac12R_2R_{1\alpha\beta}R_3^{\alpha\beta}
-\frac14R_1R_2R_3
\Big].                                       \label{eqn:6.26}
\end{eqnarray}
\arraycolsep=3pt
\mathindent=\leftmargini
Since these expressions contain arbitrary form
factors, they suggest that in virtue of (\ref{eqn:6.25a}) -
(\ref{eqn:6.25}),
in four dimensions,
the basis of nonlocal gravitational invariants
may be redundant. However, to convert
relations (\ref{eqn:6.25a}), (\ref{eqn:6.25}) into constraints
between
the basis structures one must make one more step:
eliminate the Riemann tensor.

As discussed in Sect.II, the Riemann tensor
expresses through the Ricci tensor in a nonlocal
way once the boundary conditions for the gravitational
field are specified. Elimination of the Riemann tensor from the
expressions (\ref{eqn:6.26a}) and (\ref{eqn:6.26}) with the use of
(\ref{eqn:2.2}) (to lowest order) brings them to the following form:
\mathindent=0pt
\arraycolsep=0pt
\begin{eqnarray}&&
\widetilde{\cal F}(\Box_1,\Box_2,\Box_3)
R_{1[\alpha\beta}{}^{\gamma\delta}
R_{2\gamma\delta}{}^{\kappa\mu}
R_{3\kappa]\mu}{}^{\alpha\beta}
\equiv
-\frac1{15}\left\{
\frac18
\widetilde{\cal F}'
\left(\frac{\Box_1}{\Box_2\Box_3}+\frac{\Box_2}{\Box_1\Box_3}
+\frac{\Box_3}{\Box_1\Box_2}\right)
   {R_1 R_2 R_3}\right.
\nonumber\\&&\ \
+\widetilde{\cal F}'
\left(
-\frac1{\Box_1}
-\frac1{\Box_2}
-\frac1{\Box_3}
+\frac12\frac{\Box_2}{\Box_1\Box_3}
+\frac12\frac{\Box_1}{\Box_2\Box_3}
+\frac12\frac{\Box_3}{\Box_1\Box_2}
\right)
   {R_{1\,\alpha}^\mu R_{2\,\beta}^{\alpha} R_{3\,\mu}^\beta}
\nonumber\\ &&\ \
+\frac14
\Big(\frac1{\Box_1}+\frac1{\Box_2}
 -\frac{\Box_3}{\Box_1\Box_2}\Big)
\left[\widetilde{\cal F}'
           +\widetilde{\cal F}'{\big|}_{2\leftrightarrow 3}
+\widetilde{\cal F}'
{\big|}_{1\leftrightarrow 3}
\right]   {R_1^{\mu\nu}R_{2\,\mu\nu}R_3}
\nonumber\\&&\ \
+\frac14\Big(\frac1{\Box_1\Box_2}
 +\frac1{\Box_1\Box_3}+\frac3{\Box_2\Box_3}\Big)
\left[
    -\widetilde{\cal F}'
    -\widetilde{\cal F}'
                    {\big|}_{1\leftrightarrow 2}
    -\widetilde{\cal F}'
                    {\big|}_{1\leftrightarrow 3}
\right]  {R_1^{\alpha\beta}
\nabla_\alpha R_2 \nabla_\beta R_3}
\nonumber\\ &&\ \
+\frac1{\Box_1\Box_2}\left[
  \widetilde{\cal F}'
           +\widetilde{\cal F}'{\big|}_{2\leftrightarrow 3}
  +\widetilde{\cal F}'{\big|}_{1\leftrightarrow 3}
\right]  {\nabla^\mu R_1^{\nu\alpha}
\nabla_\nu R_{2\,\mu\alpha}R_3}
\nonumber\\&&\ \
+\frac1{\Box_2\Box_3}\left[
    \widetilde{\cal F}'+\widetilde{\cal F}'
          {\big|}_{1\leftrightarrow 2}
+\widetilde{\cal F}'
                    {\big|}_{1\leftrightarrow 3}
\right]  {R_1^{\mu\nu}\nabla_\mu R_2^{\alpha\beta}\nabla_\nu
         R_{3\,\alpha\beta}}
\nonumber\\&&\ \
+\Big(\frac1{\Box_1\Box_2}+\frac1{\Box_1\Box_3}
      -\frac1{\Box_2\Box_3}\Big)\left[
    \widetilde{\cal F}'
    +\widetilde{\cal F}'{\big|}_{1\leftrightarrow 2}
    +\widetilde{\cal F}'{\big|}_{1\leftrightarrow 3}
\right]  {R_1^{\mu\nu}\nabla_\alpha R_{2\,\beta\mu}\nabla^\beta
       R_{3\,\nu}^\alpha}
\nonumber\\&&\ \
+\frac1{\Box_1\Box_2\Box_3}\left[
-\widetilde{\cal F}'
           -\widetilde{\cal F}'{\big|}_{2\leftrightarrow 3}
	   -\widetilde{\cal F}'
{\big|}_{1\leftrightarrow 3}
\right]  {\nabla_\alpha\nabla_\beta R_1^{\mu\nu}
\nabla_\mu\nabla_\nu
       R_2^{\alpha\beta} R_3}    \nonumber\\&&\ \
\left.+\frac2{\Box_1\Box_2\Box_3}\left[
-\widetilde{\cal F}'
           -\widetilde{\cal F}'{\big|}_{2\leftrightarrow 3}
	   -\widetilde{\cal F}
{\big|}_{1\leftrightarrow 3}
\right]  {\nabla_\mu R_1^{\alpha\lambda}
\nabla_\nu R_{2\,\lambda}^\beta
     \nabla_\alpha\nabla_\beta R_3^{\mu\nu}}
\right\}\nonumber\\
&&\ \ +{\rm a\ total\ derivative}+{\rm O}[R^4_{..}],\ \ \ \ \ \
\widetilde{\cal F}'\equiv(\Box_1-\Box_2-\Box_3)
\widetilde{\cal F},
\label{eqn:6.27}
\end{eqnarray}
\begin{eqnarray}
&&\widetilde{\widetilde{\cal F}}(\Box_1,\Box_2,\Box_3)
R_{1[\alpha\beta}{}^{\gamma\delta}
R_{2\gamma\delta}{}^{\alpha\beta}
R_{3\mu]}^\mu
\equiv
-\frac2{15}\left\{
\frac18\widetilde{\widetilde{\cal F}}
\left(\frac{\Box_1}{\Box_2}+\frac{\Box_2}{\Box_1}
+\frac{{\Box_3}^2}{\Box_1\Box_2}\right)
   {R_1 R_2 R_3} \right.  \nonumber\\
&&\ \
+\widetilde{\widetilde{\cal F}}\left(
-1+\frac12\frac{\Box_1}{\Box_2}+\frac12\frac{\Box_2}{\Box_1}
       -\frac{\Box_3}{\Box_1}-\frac{\Box_3}{\Box_2}
       +\frac12\frac{{\Box_3}^2}{\Box_1\Box_2}\right)
   {R_{1\,\alpha}^\mu R_{2\,\beta}^{\alpha} R_{3\,\mu}^\beta}
\nonumber\\
&&\ \
+\frac14\left[
    \widetilde{\widetilde{\cal F}}\Big(
\frac{\Box_3}{\Box_1}+\frac{\Box_3}{\Box_2}
-\frac{{\Box_3}^2}{\Box_1\Box_2}\Big)
   +{\widetilde{\widetilde{\cal F}}}
{\big|}_{1\leftrightarrow 3}
\Big(1-\frac{\Box_3}{\Box_2}+\frac{\Box_1}{\Box_2}\Big)\right.
\left.
   +{\widetilde{\widetilde{\cal F}}}
{\big|}_{2\leftrightarrow 3}
\Big(1-\frac{\Box_3}{\Box_1}+\frac{\Box_2}{\Box_1}\Big)\right]
   {R_1^{\mu\nu}R_{2\,\mu\nu}R_3}
\nonumber\\&&\ \
+\frac14\left[
    -\Big(\widetilde{\widetilde{\cal F}}
+{\widetilde{\widetilde{\cal F}}}
{\big|}_{1\leftrightarrow 2}\Big)
     \Big(\frac1{\Box_1}+\frac3{\Box_2}
+\frac{\Box_3}{\Box_1\Box_2}\Big)  \right.
\left.
    -{\widetilde{\widetilde{\cal F}}}
{\big|}_{1\leftrightarrow 3}
     \Big(\frac1{\Box_2}+\frac1{\Box_3}
+3\frac{\Box_1}{\Box_2\Box_3}\Big)
   \right]
   {R_1^{\alpha\beta}
\nabla_\alpha R_2 \nabla_\beta R_3}
\nonumber\\
&&\ \
+\left(\widetilde{\widetilde{\cal F}}
\frac{\Box_3}{\Box_1\Box_2}
+{\widetilde{\widetilde{\cal F}}}
{\big|}_{1\leftrightarrow 3}\frac1{\Box_1}
   +{\widetilde{\widetilde{\cal F}}}
{\big|}_{1\leftrightarrow 3}\frac1{\Box_2}\right)
   {\nabla^\mu R_1^{\nu\alpha}
\nabla_\nu R_{2\,\mu\alpha}R_3}
\nonumber\\&&\ \
+\left[
   {\widetilde{\widetilde{\cal F}}}
{\big|}_{1\leftrightarrow 3}
\frac{\Box_1}{\Box_2\Box_3}
   +\Big(\widetilde{\widetilde{\cal F}}
+{\widetilde{\widetilde{\cal F}}}
{\big|}_{1\leftrightarrow 2}\Big)\frac1{\Box_2}
  \right]
  {R_1^{\mu\nu}\nabla_\mu R_2^{\alpha\beta}\nabla_\nu
         R_{3\,\alpha\beta}}
\nonumber\\&&\ \
+\left[
   {\widetilde{\widetilde{\cal F}}}
{\big|}_{1\leftrightarrow 3}
\Big(\frac1{\Box_2}+\frac1{\Box_3}
-\frac{\Box_1}{\Box_2\Box_3}\Big)\right.
\left.
   +\Big(\widetilde{\widetilde{\cal F}}
+{\widetilde{\widetilde{\cal F}}}
{\big|}_{1\leftrightarrow 2}\Big)
    \Big(\frac1{\Box_1}-\frac1{\Box_2}
+\frac{\Box_3}{\Box_1\Box_2}\Big)
 \right]
  {R_1^{\mu\nu}\nabla_\alpha R_{2\,\beta\mu}\nabla^\beta
       R_{3\,\nu}^\alpha}  \nonumber\\&&\ \
+\left(-\widetilde{\widetilde{\cal F}}
\frac1{\Box_1\Box_2}
-{\widetilde{\widetilde{\cal F}}}
{\big|}_{1\leftrightarrow 3}
\frac1{\Box_2\Box_3}-{\widetilde{\widetilde{\cal F}}}
{\big|}_{2\leftrightarrow 3}\frac1{\Box_1\Box_3}\right)
  {\nabla_\alpha\nabla_\beta R_1^{\mu\nu}
\nabla_\mu\nabla_\nu
       R_2^{\alpha\beta} R_3}    \nonumber\\&&\ \
+2\left(-\widetilde{\widetilde{\cal F}}\frac1{\Box_1\Box_2}
-{\widetilde{\widetilde{\cal F}}}
{\big|}_{1\leftrightarrow 3}
\frac1{\Box_2\Box_3}-{\widetilde{\widetilde{\cal F}}}
{\big|}_{2\leftrightarrow 3}\frac1{\Box_1\Box_3}\right)
  {\nabla_\mu R_1^{\alpha\lambda} \nabla_\nu R_{2\,\lambda}^\beta
     \nabla_\alpha\nabla_\beta R_3^{\mu\nu}}
\Big\}\nonumber\\
&&\ \
+{\rm a\ total\ derivative}
+{\rm O}[R^4_{..}],                \label{eqn:6.28}
\end{eqnarray}
\arraycolsep=3pt
\mathindent=\leftmargini
where
\[
{\widetilde{\cal F}}'
{\big|}_{1\leftrightarrow 2}
\equiv\widetilde{\cal F}'
(\Box_2,\Box_1,\Box_3),\ {\rm etc}.
\]
It is now seen that, up to total derivatives and terms ${\rm
O}[R^4_{..}]$,
(\ref{eqn:6.25a}) is an equivalent form of (\ref{eqn:6.25})
corresponding to
\[
\widetilde{\widetilde{\cal F}}=
\frac12\frac1{\Box_3}\widetilde{\cal
F}'=\frac12\frac{\Box_1-\Box_2-\Box_3}{\Box_3}
\widetilde{\cal F}.
\]

Thus, assuming integration over the space-time which will
make irrelevant total derivatives, we conclude that,
at third order in the curvature, of the two generally
different identities for nonlocal invariants, existing
in four dimensions, only one is independent: eq. (\ref{eqn:6.25}).
This equation is brought to its final form by putting
\[
\widetilde{\widetilde{\cal F}}(\Box_1,\Box_2,\Box_3)
=-\frac16\Box_1\Box_2{\cal F}(\Box_1,\Box_2,\Box_3)
\]
where ${\cal F}(\Box_1,\Box_2,\Box_3)$ is a new arbitrary function,
and taking into account the symmetries of the
tensor structures entering (\ref{eqn:6.28}). The result is
the following constraint between the basis
invariants listed in the table (\ref{eqn:5.21})-(\ref{eqn:5.24}):
\begin{eqnarray}
&&
\int\! dx\, g^{1/2}\, {\rm tr}\,
{\cal F}^{\rm sym}(\Box_1,\Box_2,\Box_3)
\Big\{-\frac1{48}({\Box_1}^2+{\Box_2}^2
+{\Box_3}^2)\Re_1\Re_2\Re_3({9})
\nonumber\\
&&\ \ \ \
-\frac1{12}({\Box_1}^2+{\Box_2}^2+{\Box_3}^2
  -2{\Box_1}{\Box_2}-2{\Box_2}{\Box_3}
-2{\Box_1}{\Box_3})\Re_1\Re_2\Re_3({10})
\nonumber\\
&&\ \ \ \
-\frac18{\Box_3}({\Box_1}+{\Box_2}-{\Box_3})\Re_1\Re_2\Re_3({11})
+\frac18(3{\Box_1}+{\Box_2}+{\Box_3})\Re_1\Re_2\Re_3({22})
\nonumber\\&&\ \ \ \
-\frac12{\Box_3}\Re_1\Re_2\Re_3({23})
-\frac12{\Box_1}\Re_1\Re_2\Re_3({24})
-\frac12({\Box_2}+{\Box_3}-{\Box_1})\Re_1\Re_2\Re_3({25})
\nonumber\\
&&\ \ \ \
+\frac12\Re_1\Re_2\Re_3({27})
+\Re_1\Re_2\Re_3({28})
\Big\}+{\rm O}[\Re^4]=0,\ 2\omega=4
\label{eqn:6.29}
\end{eqnarray}
where ${\cal F}^{\rm sym}(\Box_1,\Box_2,\Box_3)$ is a {\em
completely symmetric} but otherwise arbitrary
function. This constraint, valid in four dimensions,
reduces the basis of nonlocal gravitational
invariants by one structure. With its aid one can
exclude everywhere either the structure 9 or 10 or the
{\em completely symmetric} (in the labels 1,2,3) part
of anyone of the remaining purely gravitational
structures except $\Re_1\Re_2\Re_3({29})$. The latter structure
which is the only one containing six derivatives
is absent from the constraint (\ref{eqn:6.29}) and is,
therefore, inexcludable. This can be explained by the fact
that its local version is the only independent
contraction of three Weyl tensors: eq. (\ref{eqn:6.19})
(cf. eq. (\ref{eqn:2.2})).

At least apparently, eqs. (\ref{eqn:6.25a}) and (\ref{eqn:6.25}) are
not
the most general nonlocal identities that can be written
down by antisymmetrizing five indices. More
generally, one can apply this procedure to three
tensors with
arbitrary indices and arbitrary number of uncontracted
derivatives. Therefore, to make sure that there are no
more constraints between the basis invariants, an independent
check is needed. Since, at third order in the curvature,
the maximum number of derivatives that do not contract
in the box operators is six, we begin this check
with nonlocal structures having three Ricci tensors and
six derivatives. There exist only two such, and, according to the
discussion of
Sect.IV (eq.(\ref{eqn:4.4})), only
one of them is independent:
\begin{eqnarray}
&&\hat{1}\nabla_\alpha\nabla_\beta R_1^{\gamma\delta}
   \nabla_\gamma\nabla_\delta R_2^{\mu\nu}
   \nabla_\mu\nabla_\nu R_3^{\alpha\beta}
   = \Re_1\Re_2\Re_3({29}),                      \label{eqn:6.30}\\
&&\hat{1}\nabla_\alpha\nabla_\beta R_1^{\gamma\delta}
   \nabla_\gamma\nabla_\mu R_2^{\alpha\nu}
   \nabla_\delta\nabla_\nu R_3^{\beta\mu}
= - \Re_1\Re_2\Re_3({29}) + (\dots)               \label{eqn:6.31}
\end{eqnarray}
where the ellipses $(\dots)$ stand for total derivatives and
terms with derivatives contracting in the box operators.
Eq. (\ref{eqn:6.31}) is obtained by three integrations by parts
applied to $\nabla_\alpha, \nabla_\mu$ and $\nabla_\delta$.
Since not more than one index of each Ricci tensor
and not more than one derivative acting on each Ricci tensor
may participate in the antisymmetrization (otherwise
the result will be either trivial or ${\rm O}[R^4_{..}]$), there
are only two possible 5 - antisymmetrizations of (\ref{eqn:6.30}):
\[
\nabla_{[\alpha} \nabla^\beta R_{1\delta}^\gamma
\nabla_\gamma \nabla^\delta R_{2\nu}^\mu
\nabla_{\mu]} \nabla^\nu R_3^{\alpha\beta} = 0,\;\;\;
\nabla_\alpha \nabla_\beta R_{1[\delta}^\gamma
\nabla_\gamma \nabla^\delta R_{2\nu}^\mu
\nabla_{\mu} \nabla^\nu R_{3\beta]}^{\alpha} = 0.
\]
In each of these cases, upon calculation, the terms
(\ref{eqn:6.30}) and (\ref{eqn:6.31}) appear in a sum with equal
coefficients
and, therefore, cancel. This proves that the structure
with six derivatives remains unconstrained. Among the
invariants with three Ricci tensors and four derivatives,
only two are independent: the basis structures 27 and 28,
and only the latter admits nontrivial 5-antisymmetrizations.
There is, moreover, only one such:
\[
\nabla_{[\mu} R_{1\lambda}^\alpha
\nabla^\nu R_{2\beta}^\lambda
\nabla_\alpha\nabla^\beta R^\mu_{3\nu]} = 0.
\]
Upon calculation and multiplication by an arbitrary form
factor, the latter identity gives precisely the constraint
(\ref{eqn:6.29}). The invariants with the commutator curvature and
four derivatives are all reducible except the structure
$\Re_1\Re_2\Re_3({38})$ (see Sect.V) which only admits the nontrivial
5-antisymmetrization of five indices, but one can apply to this
structure the
same argumentation as in the case of $\Re_1\Re_2\Re_3({29})$ above,
which
proves that it also remains unconstrained. Finally, invariants
with two derivatives do not admit a nontrivial
5-antisymmetrization since, for that, one needs at
least ten indices: five uncontracted to be involved in
the antisymmetrization, and five more to make a
complete contraction.

Thus, in four dimensions, there is only one constraint
between the basis structures, and the dimension of the basis
of nonlocal cubic invariants which is generally 38
and in the case of the gravitational invariants 10
becomes respectively 37 and 9.

The nonlocal identity obtained
above has a direct relation to the Gauss-Bonnet identity in four
dimensions. Indeed, by calculating the square of the Riemann tensor
with the
aid of eq. (\ref{eqn:2.2}), one
finds for arbitrary dimension $2\omega$ :
\mathindent=0pt
\arraycolsep=0pt
\begin{eqnarray}&&
\int\! dx\, g^{1/2}\, \,\Big(
R_{\alpha\beta\gamma\delta}
R^{\alpha\beta\gamma\delta}
-4 R_{\mu\nu} R^{\mu\nu} + R^2
\Big)
 =\int\! dx\, g^{1/2}\, \, \left[\frac12\frac{\Box_1}{\Box_2\Box_3}
   {R_1 R_2 R_3} \right.
\nonumber\\
&& \quad\quad
+2\Big(\frac{\Box_1}{\Box_2\Box_3}-2\frac1{\Box_1}\Big)
   {R_{1\,\alpha}^\mu R_{2\,\beta}^{\alpha}
R_{3\,\mu}^\beta}
+\Big(2\frac1{\Box_1}-\frac{\Box_3}{\Box_1\Box_2}\Big)
   {R_1^{\mu\nu}R_{2\,\mu\nu}R_3}
\nonumber\\
&& \quad\quad
+\Big(-2\frac1{\Box_1\Box_3}-3\frac1{\Box_2\Box_3}\Big)
   {R_1^{\alpha\beta}
\nabla_\alpha R_2 \nabla_\beta R_3}
+4\frac1{\Box_1\Box_2}
   {\nabla^\mu R_1^{\nu\alpha}
\nabla_\nu R_{2\,\mu\alpha}R_3}
\nonumber\\
&& \quad\quad
+4\frac1{\Box_2\Box_3}
  {R_1^{\mu\nu}\nabla_\mu R_2^{\alpha\beta}
\nabla_\nu
         R_{3\,\alpha\beta}}
+4\Big(2\frac1{\Box_1\Box_2}-\frac1{\Box_2\Box_3}\Big)
  {R_1^{\mu\nu}
\nabla_\alpha R_{2\,\beta\mu}\nabla^\beta
       R_{3\,\nu}^\alpha}  \nonumber\\
&& \quad\quad
-4\frac1{\Box_1\Box_2\Box_3}
  {\nabla_\alpha\nabla_\beta R_1^{\mu\nu}
\nabla_\mu\nabla_\nu
       R_2^{\alpha\beta} R_3}
-8\frac1{\Box_1\Box_2\Box_3}
{\nabla_\mu R_1^{\alpha\lambda}
\nabla_\nu R_{2\,\lambda}^\beta
\nabla_\alpha\nabla_\beta R_3^{\mu\nu}}\Big]
+{\rm O}[R^4_{..}].                              \label{eqn:6.33}
\end{eqnarray}
\arraycolsep=3pt
\mathindent=\leftmargini
In agreement with the result of Ref.\onlinecite{II}, a contribution
of second order in the curvature is absent from this expression
for any space-time dimension. The third-order contribution
(\ref{eqn:6.33}) does not generally vanish but vanishes in four
dimensions because it coincides with the left-hand side of
the identity (\ref{eqn:6.29}) if in the latter one puts
\[
{\cal F}^{\rm sym}(\Box_1,\Box_2,\Box_3) =
-8({\rm tr}\hat{1})^{-1}\frac1{\Box_1\Box_2\Box_3}.
\]
Comparison of eq.(\ref{eqn:6.33}) and (\ref{eqn:6.28}) gives
\begin{eqnarray}&&
\int\! dx\, g^{1/2}\, \,\Big(
R_{\alpha\beta\gamma\delta}
R^{\alpha\beta\gamma\delta}
-4 R_{\mu\nu} R^{\mu\nu} + R^2
\Big) \nonumber\\
&&\quad\quad\quad\quad\quad\ \ \ \ \ \ \ \
=\int\! dx\, g^{1/2}\, \,
\Big(-10\frac1{\Box_3}\Big)
R_{1[\alpha\beta}{}^{\gamma\delta}
R_{2\gamma\delta}{}^{\alpha\beta}
R_{3\mu]}^\mu +
{\rm O}[R^4_{..}].                            \label{eqn:6.34}
\end{eqnarray}
This relation {\em valid for any number of space-time dimensions}
elucidates the mechanism by which the Gauss-Bonnet
identity arises in four dimensions \cite{HOsborn}.

\section*{Acknowledgments}
\indent
The work of A.O.B. on this paper was partially supported by the CITA
National Fellowship and NSERC grant at the University of Alberta.
G.A.V. has been supported by the Russian Science Foundation under
Grant
93-02-15594 and by a NATO travel grant (CRG 920991). V.V.Z.,
whose work was supported in part by National Science Council of the
Republic of China under contract No. NSC 82-0208-M008-070, thanks the
Department of Physics of National Central University for the support
in
computing facilities.


\begin{thebibliography}{99}

\bibitem{DW63} B.S.DeWitt, in {\it Relativity, groups and topology},
1963 Les
Houches lectures, eds. C.DeWitt and B.S.DeWitt (Gordon and Breach,
N.Y., 1964)
p.587; {\it Dynamical theory of groups and fields }
(Gordon and Breach, New York, 1965)

\bibitem{JLas} G.Jona-Lasinio, Nuovo Cim. {\bf 34}, 1790 (1964)

\bibitem{Hooft-Hon}J.Honerkamp, Nucl.Phys. {\bf B 36}, 130 (1972);
G.t'Hooft,
in Proc. XII Winter School in Carpacz, Acta Universitatis
Wratislaviensis,
No.368 (1975)

\bibitem{DDI} S.Deser, M.Duff and C.J.Isham, Nucl.Phys. {\bf B111},
45 (1976)

\bibitem{Gosp} G.A.Vilkovisky, in {\it Quantum theory of gravity},
ed.
S.M.Christensen (Hilger, Bristol, 1984) p.169

\bibitem{in}J.Schwinger, J.Math.Phys. {\bf 2}, 407 (1961);
L.V.Keldysh,
Zh.Eksp.Teor.Fiz. {\bf 47}, 1515 (1964); Yu.V.Golfand, Yad.Fiz. {\bf
8}, 600
(1968); P.Hajicek, in Proc. Second Marcel Grossman meeting on general
relativity, Trieste, 1979, ed. R.Ruffini (North Holland, 1982) p.483;
E.S.Fradkin and D.M.Gitman, Fortschr. der Phys. {\bf 29}, 381 (1981);
J.L.Buchbinder, E.S.Fradkin and D.M.Gitman, Fortschr. der Phys. {\bf
29}, 187
(1981); R.D.Jordan, Phys.Rev. {\bf D33}, 44 (1986); E.Calzetta and
B.L.Hu,
Phys.Rev. {\bf D35}, 495 (1987)

\bibitem{Frolov} V.P.Frolov and G.A.Vilkovisky, Proc. Second Seminar
on Quantum
Gravity (Moscow, 1981), eds. M.A.Markov and P.C.West (Plenum, London,
1983)
p.267; Phys.Lett. {\bf 106 B}, 307 (1981)

\bibitem{Hartle} J.B.Hartle and G.Horowitz, Phys.Rev. {\bf D24}, 257
(1981)

\bibitem{PRep}
A.O.Barvinsky and G.A.Vilkovisky,
Phys. Reports {\bf 119}, 1 (1985)

\bibitem{Frfest}
A.O.Barvinsky and G.A.Vilkovisky, in
{\it Quantum field theory and quantum statistics},
vol. 1, eds. I.A.Batalin, C.J.Isham and
G.A.Vilkovisky
(Hilger, Bristol, 1987) p.245

\bibitem{I}
A.O.Barvinsky and G.A.Vilkovisky,
Nucl.Phys. {\bf B282}, 163 (1987)

\bibitem{II}
A.O.Barvinsky and G.A.Vilkovisky,
Nucl. Phys. {\bf B333}, 471 (1990)

\bibitem{III}A.O.Barvinsky and G.A.Vilkovisky,
Nucl.Phys. {\bf B333}, 512 (1990)

\bibitem{IV}A.O.Barvinsky, Yu.V.Gusev, G.A.Vilkovisky
and V.V.Zhytnikov,
Covariant Perturbation Theory (IV). Third Order in the Curvature,
Report of the
University of Manitoba (University of Manitoba, Winnipeg, 1993)

\bibitem{CQG}G.A.Vilkovisky, Class.Quantum Grav. {\bf 9}, 895 (1992)

\bibitem{Armen}
A.G.Mirzabekian and G.A.Vilkovisky, Phys.Lett. {\bf B317}, 517 (1993)

\bibitem{DW83}B.S.DeWitt, in {\it Relativity, groups and topology},
1963 Les
Houches lectures, eds. B.S.DeWitt and R.Stora (North Holland, 1984)
p.221

\bibitem{Hawking}S.W.Hawking, Commun.Math.Phys. {\bf 43}, 199 (1975)

\bibitem{GAV-Strasb}
G.A.Vilkovisky, Preprint CERN-TH.6392/92; Publication de l'Institut
de
Recherche Math\'ematique Avanc\'ee, R.C.P. 25, vol.43 (Strasbourg,
1992) p.203

\bibitem{Schw2} J.S.Schwinger, Phys.Rev. {\bf 82}, 664 (1951)

\bibitem{Zeln}
A.I.Zelnikov,
Phys.Lett. {\bf B273}, 471 (1991);
A.V.Leonidov and A.I.Zelnikov,
Phys.Lett. {\bf B276}, 122 (1992)

\bibitem{JMP} A.O.Barvinsky, Yu.V.Gusev, G.A.Vilkovisky
and V.V.Zhytnikov,
Asymptotic behaviours of the heat kernel in covariant perturbation
theory, in
press

\bibitem{freaks}A special status of structures (\ref{eqn:5.25}) is
not
exceptional in field-theoretic calculations. For example, the
first-order
invariant $R$ is also absent in the expansion (\ref{eqn:5.26}) for
the heat
kernel trace (although it usually emerges in such an expansion as a
part of the
invariant ${\rm tr}\,\hat P$), while the second-order structure
$\Re_1\Re_2({3})$ has a peculiarity that its one-loop effective
action form
factor is a local numerical constant \cite{II}.

\bibitem{Fr-Zeln}
V.P.Frolov and A.I.Zelnikov, Phys.Rev. {\bf D29}, 1057 (1984)

\bibitem{26}
S.A.Fulling, R.C.King, B.G.Wybourne
and C.J.Cummins,
Class. Quantum Grav. {\bf 9}, 1151 (1992)

\bibitem{DesSch}S.Deser and A.Schwimmer, Preprint Brandeis BRX-343,
SISSA
14/93/EP

\bibitem{HOsborn} In Ref.\onlinecite{IV} this equation figures with
the Weyl
tensors instead of the Riemann ones -- the fact brought to our
attention by
Hugo Osborn, and the authors are very grateful to him for pointing
\nopagebreak
out this error.

\end{thebibliography}
\end{document}